\documentclass[journal,final,twocolumn,10pt,twoside]{IEEEtranTCOM}

\normalsize

\usepackage{mathtools}
\usepackage{amsmath}
\usepackage{amssymb}
\usepackage{amsmath}
\usepackage{cite}
\usepackage{algorithm}
\usepackage{enumerate,multirow}
\usepackage{algpseudocode}

\usepackage{siunitx}
\usepackage{multirow}
\usepackage{multicol}
\usepackage{color}
\usepackage{xcolor}

\usepackage{graphicx}
\usepackage{graphicx}
\date{}

\usepackage{times,amssymb,amsmath,amsfonts,nicefrac,color,bbm,mathrsfs,float}
\usepackage{multirow,tikz,graphicx}

\usepackage{xcolor}
\usepackage{comment}
\definecolor{MyGreen1}{RGB}{20,180,40}
\definecolor{MyBlue1}{RGB}{00,150,255}
\definecolor{MyGray1}{RGB}{200,200,200}

\usetikzlibrary{shapes.misc, positioning}
\usetikzlibrary{decorations.pathreplacing}

\usepackage{makecell}
\usetikzlibrary{fit}

\usetikzlibrary{arrows.meta,
	chains,
	decorations.pathreplacing,calligraphy,
	positioning,
	quotes,
	shapes.geometric,
	decorations.text
}

\tikzset{
	BC/.style = {decorate,  
		decoration={calligraphic brace, amplitude=1.2mm,
			raise=1.2mm, mirror},
		thick, pen colour={black}
	},
}

\tikzset{
	BC2/.style = {decorate,  
		decoration={calligraphic brace, amplitude=2mm,
			raise=1.5mm, mirror},
		thick, pen colour={black}
	},
}

\tikzset{
	block/.style    = {draw, thick, rectangle, text centered, align=center,  minimum width = 2em},
	sblock/.style      = {draw, thick, rectangle, minimum height = 2em,
		minimum width = 2em}, 
}

\tikzstyle{block} = [rectangle, draw,  text centered]
\tikzstyle{block} = [rounded corners]

\usepackage{calc}
\newcolumntype{M}[1]{>{\centering\arraybackslash}m{#1}}
\newcolumntype{N}{@{}m{0pt}@{}}

\usepackage{mdwmath}
\usepackage{blindtext}
\usepackage{eqparbox}
\usepackage{fixltx2e}
\usepackage{stfloats}
\DeclareMathOperator{\E}{\mathbb{E}}

\usepackage{amsthm}

\DeclareMathAlphabet{\mathbfsl}{OT1}{ppl}{b}{it} 

\newcommand{\ceil}[1]{\left\lceil #1 \right\rceil}

\newcommand{\be}[1]{\begin{equation}\label{#1}}
	\newcommand{\ee}{\end{equation}}

\include{comments}

\newcommand{\Cref}[1]{Co\-ro\-lla\-ry\,\ref{#1}}

\begin{document}
\title{ProductAE: Toward Deep Learning Driven Error-Correction Codes of Large Dimensions}
	\author{Mohammad Vahid Jamali, Hamid Saber, Homayoon Hatami, and Jung Hyun Bae
		\vspace{-0.4cm}
		\thanks{This paper was presented in part at the IEEE International Conference on Communications (ICC), Seoul, South Korea, May 2022 \cite{jamali2022productae}.}
		\thanks{The authors are with the SOC Lab at Samsung Semiconductor, Inc., San Diego, CA. Emails: \{mv.jamali, hamid.saber, h.hatami, jung.b\}@samsung.com.}
	}
\maketitle

\begin{abstract}
While decades of theoretical research have led to the invention of several classes of error-correction codes, the design of such codes is an extremely challenging task, mostly driven by human ingenuity. Recent studies demonstrate that such designs can be effectively automated and accelerated via tools from machine learning (ML), thus enabling 
 ML-driven classes of error-correction codes with promising performance gains compared to classical designs. A fundamental challenge, however, is that it is prohibitively complex, if not impossible, to design and train fully ML-driven encoder and decoder pairs for large code dimensions.
In this paper, we propose Product Autoencoder (ProductAE) -- a computationally-efficient family of deep learning driven (encoder, decoder) pairs -- aimed at enabling the training of relatively large 
 codes (both encoder and decoder) with a manageable training complexity. We build upon ideas from classical product codes and propose constructing large neural codes using smaller code components. ProductAE boils down the complex problem of training the encoder and decoder for a large code dimension $k$ and blocklength $n$ to less-complex sub-problems of training encoders and decoders for smaller dimensions and blocklengths.
Our training results show successful training of ProductAEs of dimensions as large as $k=300$ bits with 
meaningful
 performance gains compared to state-of-the-art classical and neural designs. Moreover, we demonstrate excellent robustness and adaptivity of ProductAEs to channel models different than the ones used for training.
\end{abstract} 

\begin{keywords} 
Product autoencoder (ProductAE), deep learning, 
error-correction 
coding, product codes, neural networks.
\end{keywords}
\section{Introduction}\label{Sec1}
\IEEEPARstart{E}RROR-CORRECTION CODING is a principal technique used for reliable information processing in the presence of unavoidable random errors. 
In practical communication systems, channel coding is used as an essential building block
to enable reliable communication by protecting the transmission of messages across a random noisy channel. 
In short, the channel encoder maps the length-$k$ sequence of information bits $\mathbf{u}$ to a length-$n$ sequence of coded symbols $\mathbf{c}$ by adding some sort of redundancy. The decoder, on the other hand, exploits this redundancy to map the noisy observations of codewords $\mathbf{y}$ back to information sequences $\mathbf{u}$ while minimizing the error rate. The parameters $k$ and $n$ as defined above are referred to as the code dimension and blocklength, respectively, and the resulting code is denoted by an $(n,k)$ code.

Channel coding is one of the most fundamental problems in communication theory \cite{shannon1948mathematical,shannon1949communication}, 
and decades of extensive theoretical research have led to the invention of several landmark codes, such as Turbo codes \cite{berrou1993near}, low-density parity-check (LDPC) codes \cite{gallager1962low}, and polar codes \cite{arikan2009channel}, among others.
The design of such classical codes, however,
is an extremely difficult task and hugely relies on human intelligence, thus slowing down new discoveries in the design of efficient encoders and decoders. 

With the success of artificial intelligence (AI) 
in many different domains, recently, there has been tremendous interest in the coding theory community to automate and accelerate the design of channel encoders and decoders by incorporating 
various tools from machine learning (ML) \cite{kim2020physical,kim2018deepcode,jiang2019turbo,o2017introduction,makkuva2021ko,o2016learning,jamali2021Reed,jamali2023machine, ye2019circular,nachmani2016learning, gruber2017deep, nachmani2018deep, vasic2018learning, teng2019low, buchberger2020prunin,xu2017improved , cammerer2017scaling, bennatan2018deep, doan2018neural, ebada2019deep, devroye2022interpreting, chen2022improving,  kim2023learning}. This is typically done by replacing the encoder and decoder (or some components within the encoder and decoder architectures) with neural networks or some trainable models. 
The objectives and gains are vast, such as reducing the encoding and decoding complexity, improving upon the performance of classical channel codes, applications to realistic 
channel models
and to emerging use cases, and designing universal decoders that simultaneously decode several codes, to mention but a few. 

A major technical challenge here is the dimensionality issue that arises in the channel coding context due to huge code spaces; there are $2^k$ distinct codewords for a binary linear code of dimension $k$. In fact, only a small fraction of all codewords will be seen during the training phase, and thus the trained models for the encoders and decoders may fail in generalizing to unseen codewords, which constitute the major fraction of codewords for a relatively large $k$.\footnote{Note that the all-zero codeword assumption that is typically used in traditional coding is not applicable when we consider general neural codes.} Additionally, a straightforward design of neural encoders and decoders for large code dimensions and lengths requires using huge networks with excessively large number of trainable parameters. These all together make it prohibitively complex to design and train relatively large neural channel encoders and decoders.
 Consequently, the current literature is mostly limited to relatively short codes having less than $k=100$ bits of information.

Another major challenge is the joint training of the encoder and decoder due to local optima that may occur as a result of non-convex  loss functions. As such, most of the literature focuses on decoding well-known codes using data-driven neural decoders \cite{nachmani2016learning, gruber2017deep, nachmani2018deep, vasic2018learning, teng2019low, buchberger2020prunin,xu2017improved, cammerer2017scaling, bennatan2018deep, doan2018neural}, and only a handful of studies focus on discovering both encoders and decoders \cite{jiang2019turbo,o2016learning, o2017introduction,makkuva2021ko}.
It is worth mentioning that the pair of encoder and decoder together can be viewed as an over-complete autoencoder, 
where the goal is to find a higher-dimensional representation of the input data  such that the original data can be reliably recovered from a noisy version of the higher-dimensional representation \cite{jiang2019turbo}.

 \begin{figure*}[t]
	\centering
	\includegraphics[trim=2cm 12.4cm 2cm 12.7cm,width=6.8in]{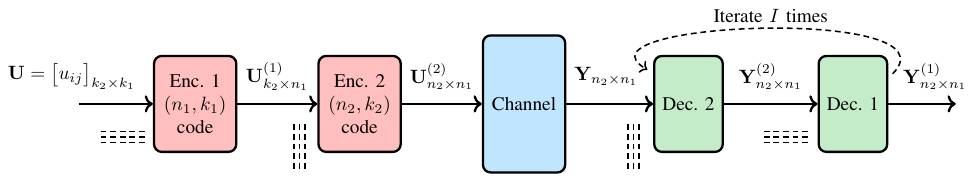}
	\vspace{-0.2cm}
	\caption{
		Block diagram of a typical communication system incorporating two-dimensional (2D) product codes. Enc. 1 and Dec. 1 encode and decode rows of the 2D input arrays while Enc. 2 and Dec. 2 encode and decode the columns of their input arrays.}
	\label{fig1}
	\vspace{-0.1cm}
\end{figure*}

In this paper, we introduce the product autoencoder (ProductAE) -- a new class of neural channel codes, i.e., a family of deep learning driven (encoder, decoder) pairs -- with two major contributions: First, it is a pioneering work that targets enabling the training of large channel codes; and second, it is the first work on designing neural product codes.
To this end, we build upon  ideas from classical product codes, and propose constructing large neural codes using smaller code components. Particularly, instead of directly training the encoder and decoder for a large code of dimension $k$ and blocklength $n$, we provide a framework that requires training smaller neural encoder and decoder components of  parameters $(n_1,k_1),(n_2,k_2), \cdots, (n_M,k_M)$ such that
$n_1n_2…n_M=n$ and $k_1k_2…k_M=k$ 
for some positive integer $M$. In other words, ProductAE boils down the complex problem of training autoencoders of large dimensions and lengths to less-complex sub-problems of training encoders and decoders for smaller dimensions and lengths.

Our training results for a relatively short-length ProductAE of 
dimension $k=100$
 show significant performance gains compared to 
 the polar code under successive cancellation (SC) decoding. More importantly, we demonstrate achieving  similar gains for moderate-length ProductAEs of 
dimensions as large as $k=300$ bits.
 This clearly establishes the generalization of our proposed architecture for training higher-dimension codes. 
Additionally, our results demonstrate meaningful gains over Turbo Autoencoder (TurboAE) \cite{jiang2019turbo} and state-of-the-art classical codes. These achievements are attained by applying innovative ideas from deep learning and intuitions from coding theory to further improve the training performance of our proposed ProductAE architecture.

The main contributions of the paper are summarized as follows.
\begin{itemize}
	\item We introduce a new class of neural error-correction codes, namely ProductAE, aimed at enabling the training of higher-dimension codes. Building upon ideas from classical product codes, ProductAE boils down the complex problem of training large autoencoders to less-complex sub-problems of training smaller encoding and decoding components.
	\item We present several useful modifications to our proposed ProductAE architecture, which significantly contribute to achieving  excellent training performances. These include heterogeneous designs, channel output injection, adding feature size, and subtracting soft information, to be detailed in Section \ref{modif}.
	\item 
	 We demonstrate the robustness and adaptivity of ProductAEs to channel models different than the ones used for training the ProductAE models. 
	\item We provide extensive sets of training examples to study the performance of ProductAEs of different parameters and to explore the usefulness of various modifications. We also investigate the impact of several additional design and training approaches. 
\end{itemize}

The rest of the paper is organized as follows. In Section \ref{prelim}, we briefly present the preliminaries in channel coding and classical product codes. In Section \ref{ProductAE}, we present the proposed ProductAE architecture together with several training and design strategies essential to achieving excellent training results. Section \ref{experiments} provides additional implementation details and presents extensive sets of training experiments to thoroughly investigate the ProductAE performance. Section \ref{nonAWGN} explores the robustness and adaptivity of ProductAEs to channel models different than the ones used for training, and Section \ref{conc} concludes the paper.
Additional design and training investigations are also provided in Appendix \ref{app1}.

\section{Preliminaries}\label{prelim}
\subsection{Channel Coding}\label{prob_form}
Consider  transmission of a length-$k$ sequence of information bits $\mathbf{u}$ across a noisy channel. The channel encoder $\mathcal{E}(\cdot)$ maps $\mathbf{u}$ to a length-$n$ sequence of coded symbols $\mathbf{c}=\mathcal{E}(\mathbf{u})$ called a codeword. Here, $k$ and $n$ are the code dimension and blocklength, respectively, and the resulting code is denoted by an $(n,k)$ code. Also, the code rate is defined as $R=k/n$.
In the case of transmission over the additive white Gaussian noise (AWGN) channel,
  the received noisy signal can be expressed as $\mathbf{y}=\mathbf{c}+\mathbf{n}$, where $\mathbf{n}$ is the channel noise vector (independent from $\mathbf{c}$) whose components are Gaussian random variables with mean zero and variance $\sigma^2$. The ratio of the average energy per coded symbol to the noise variance is called the signal-to-noise ratio (SNR). In this paper,
we assume that the encoder satisfies a soft power constraint such that the average power per coded bit is equal to $1$ (see Eq. \eqref{norm}), and 
we define the SNR as
 ${\rm SNR}=1/\sigma^2$.

The decoder $\mathcal{D}(\cdot)$, on the other hand, exploits the added redundancy by the encoder to estimate the transmitted messages from the noisy received sequences as $\hat{\mathbf{u}}=\mathcal{D}(\mathbf{y})$. The main objective of channel coding is to increase the reliability by minimizing the error rate that is often measured by the 
bit error rate (BER) or block error rate (BLER), defined as ${\rm BER}=\frac{1}{k}\sum_{i=1}^k{\Pr}(\hat u_i \neq u_i)$ and ${\rm BLER}=\Pr(\hat{\mathbf{u}} \neq \mathbf{u})$, respectively, where $x_i$ denotes the $i$-th element of a vector $\mathbf{x}$.

\subsection{Product Codes}\label{PCs}
Product coding  is a powerful coding scheme in constructing large error-correction codes. Building large codes upon product codes can provide several advantages such as low encoding and decoding complexity, large minimum distances, and a highly parallelizable implementation \cite{elias1954error,pyndiah1998near,mukhtar2016turbo,jamali2022low}. 
The encoding and decoding procedure of two-dimensional (2D) product codes is schematically shown in Fig. \ref{fig1}. Considering two codes $\mathcal{C}_1:(n_1,k_1)$ and $\mathcal{C}_2:(n_2,k_2)$, their product code,
that has the length $n=n_1 n_2$ and dimension $k=k_1 k_2$,
 is constructed in the following steps:
 \begin{enumerate}
 	\item Reshaping the length-$k_1 k_2$ information sequence as a $k_2\times k_1$ matrix $\mathbf{U}$.
 	\item Encoding each row of $\mathbf{U}$ using $\mathcal{C}_1$ to get a matrix $\mathbf{U}^{(1)}$ of size $k_2\times n_1$ (or encoding each column using $\mathcal{C}_2$ to get a matrix of size $n_2\times k_1$)
 	\item Encoding each column of $\mathbf{U}^{(1)}$ using $\mathcal{C}_2$ to get a matrix $\mathbf{U}^{(2)}$ of size $n_2\times n_1$ (or encoding each row using $\mathcal{C}_1$ to get a matrix of size $n_2\times n_1$).
 \end{enumerate}
The encoded matrices $\mathbf{U}^{(2)}$ can then be reshaped to vectors $\mathbf{c}$ of length $n=n_2\times n_1$, which represent the codewords of the 2D product code. Upon the transmission of the codewords across the noisy channel, the reshaped noisy matrices $\mathbf{Y}_{n_2\times n_1}$ are formed, from which the decoding proceeds in similar steps to encoding by first decoding the columns and then the rows (or vice versa) of the received noisy signals.

In general, to construct an $M$-dimensional product code $\mathcal{C}$, one needs to iterate $M$ codes $\mathcal{C}_1,\mathcal{C}_2,\cdots,\mathcal{C}_M$. In this case, each $m$-th encoder, $m=1,\cdots M$, encodes the $m$-th dimension of the $M$-dimensional input array.
Similarly, each $m$-th decoder decodes the noisy vectors in the $m$-th dimension of the received noisy signal.
Assuming $\mathcal{C}_m:(n_m,k_m,d_m)$ with the generator matrix $\mathbf{G}_m$, where $d$ stands for the minimum distance, the parameters of the resulting product code $\mathcal{C}$ can be obtained as the product of the parameters of the component codes, i.e.,
\begin{align}
x&=\prod_{m=1}^{M}x_m,\hspace{1cm} x\in\{n,k,d,R\},\label{X}\\
	\mathbf{G}&=\mathbf{G}_1\otimes \mathbf{G}_2\otimes\cdots\otimes\mathbf{G}_M,\label{G}
\end{align}
where $\otimes$ is the Kronecker product operator.
A near-optimum iterative decoding algorithm for 
the product of linear block codes was
proposed in \cite{pyndiah1998near}, which is based on soft-input soft-output (SISO) decoding of the component codes. It has been shown in \cite{pyndiah1998near} and \cite{mukhtar2016turbo} that
SISO decoding together with applying several decoding iterations deliver great decoding performances for product codes.
In other words, errors that have not been corrected over column decoding may be correctly decoded over row decoding or in the next decoding iterations.
Interested readers are referred to \cite{mukhtar2016turbo} and the references therein for more details on product codes.  


\begin{algorithm}[t]
	\caption{ProductAE Training Algorithm}    \label{alg_PAE_training}
	\textbf{Input:} 
	Dimension of ProductAE $M$, 
	Set of code dimensions $\mathcal{K}:=\{k_1,k_2,\cdots,k_M\}$ and blocklengths $\mathcal{N}:=\{n_1,n_2,\cdots,n_M\}$,
	number of epochs $E$, batch size $B$, number of encoder and decoder training iterations $T_{\rm enc}$ and  $T_{\rm dec}$, encoder training SNR $\gamma_{\rm e}$, decoder training SNR range $[\gamma_{\rm d,l},\gamma_{\rm d,u}]$, and encoder and decoder learning rates ${\rm lr_{enc}}$ and ${\rm lr_{dec}}$.\\	
	\textbf{Output:} Encoder and decoder NN weights $\mathbf{\Phi}$ and $\mathbf{\Theta}$.
	\vspace*{0.05in}
	\begin{algorithmic}[1]
		\State Initialize $\mathbf{\Phi}$ and $\mathbf{\Theta}$
		\For {$e=1,\cdots,E$} \Comment perform $E$ training epochs
		\For {$i_{\rm d}=1,\cdots,T_{\rm dec}$} \Comment decoder training schedule
		\State $\mathbf{U}\gets$ generate a batch of $B$ message words 
		\State $\mathbf{C}\gets\texttt{ProductAE}\_\texttt{Enc}(\mathbf{U},\mathbf{\Phi},\mathcal{K},\mathcal{N})$
		\State $\mathbf{N}\gets$ generate a batch of $B$ noise vectors with SNRs from the range $[\gamma_{\rm d,l},\gamma_{\rm d,u}]$
		\State $\mathbf{Y}\gets\mathbf{C}+\mathbf{N}$ \Comment batch of noisy codewords
		\State $\hat{\mathbf{U}}\gets\texttt{ProductAE}\_\texttt{Dec}(\mathbf{Y},\mathbf{\Theta},\mathcal{K},\mathcal{N})$
		\State $\mathbf{\Theta}\gets\texttt{Optimizer}(\mathcal{L}(\mathbf{U},\hat{\mathbf{U}}),\mathbf{\Theta},{\rm lr_{dec}})$ \Comment apply the optimizer to update $\mathbf{\Theta}$ while keeping $\mathbf{\Phi}$ fixed
		\EndFor
		\For {$i_{\rm e}=1,\cdots,T_{\rm enc}$} \Comment encoder training schedule
		\State $\mathbf{U}\gets$ generate a batch of $B$ message words 
		\State $\mathbf{C}\gets\texttt{ProductAE}\_\texttt{Enc}(\mathbf{U},\mathbf{\Phi},\mathcal{K},\mathcal{N})$
		\State $\mathbf{N}\gets$ a batch of $B$ noise vectors with SNR $\gamma_{\rm e}$
		\State $\mathbf{Y}\gets\mathbf{C}+\mathbf{N}$ 
		\State $\hat{\mathbf{U}}\gets\texttt{ProductAE}\_\texttt{Dec}(\mathbf{Y},\mathbf{\Theta},\mathcal{K},\mathcal{N})$
		\State $\mathbf{\Phi}\gets\texttt{Optimizer}(\mathcal{L}(\mathbf{U},\hat{\mathbf{U}}),\mathbf{\Phi},{\rm lr_{enc}})$ \Comment apply the optimizer to update $\mathbf{\Phi}$ while keeping $\mathbf{\Theta}$ fixed
		\EndFor
		\EndFor
		\State \textbf{return} $\mathbf{\Phi}$ and $\mathbf{\Theta}$
	\end{algorithmic}
\end{algorithm}

\begin{figure*}[t]
	\centering
	\includegraphics[trim=2cm 10cm 2cm 10.7cm,width=6.8in]{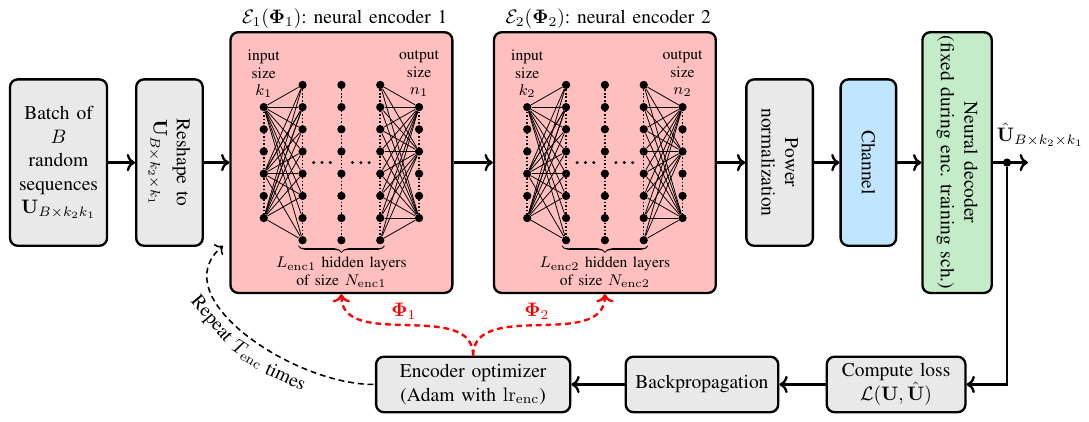}
	\vspace{-0.4cm}
	\caption{Encoder training schedule. The architecture of the 2D (i.e., $M=2$) ProductAE encoder is also depicted.}
	\label{fig2}	
	\vspace{-0.1cm}
\end{figure*}

\section{Product Autoencoder}\label{ProductAE}
\subsection{Proposed Architecture and Training}\label{proposed}
In this paper, we build upon 2D product codes. However, the proposed architectures and the methodologies naturally apply to higher dimensions. It is worth mentioning that the choice of 2D product codes balances a trade-off between the  complexity and performance. Indeed, from the classical coding theory perspective, although higher-dimensional product codes have lower encoding and decoding complexity for a given overall code size, they may result in inferior performance compared to lower-dimensional product codes due to exploiting the product structure to a greater extent (which is suboptimal compared to a direct design for the large code size).

In our ProductAE architecture, each encoder is replaced by a distinct neural network (NN). Also, assuming $I$ decoding iterations, each pair of decoders at each iteration is replaced by a distinct pair of decoding NNs resulting in $2I$ decoding NNs in total. Given that the ProductAE structure enables training encoders and decoders with relatively small code components, fully-connected NNs (FCNNs) are applied throughout the paper\footnote{We also explored convolutional NN (CNN)-based ProductAEs. However, we were able to achieve better training results with the FCNN-based implementation.}.

As summarized in Algorithm \ref{alg_PAE_training}, ProductAE training is comprised of two main schedules: $(i)$ decoder training schedule; and $(ii)$ encoder training schedule. More specifically, at each training epoch, we first train the decoder several times while keeping the encoder network fixed, and then we train the encoder multiple times while keeping the decoder weights unchanged. In the following, we first present the encoder architecture and describe its training schedule, and then we
present the decoder architecture and training schedule.

\begin{algorithm}[t]
	\caption{ProductAE Encoder Function $\texttt{ProductAE}\_\texttt{Enc}$}    \label{alg_PAE_Enc}
	\textbf{Input:} Batch of input information sequences $\mathbf{U}$,
	set of code dimensions $\mathcal{K}:=\{k_1,k_2,\cdots,k_M\}$ and blocklengths $\mathcal{N}:=\{n_1,n_2,\cdots,n_M\}$,
	and encoder NNs\\
	\textbf{Output:} Batch of codewords $\mathbf{C}$.
	\vspace*{0.05in}
	\begin{algorithmic}[1]
		\State $\mathbf{U}\gets$ reshape each sequence of bits in $\mathbf{U}$ to an $M$-dimensional array according to the dimensions set $\mathcal{K}$
		\For {$m=1,\cdots,M$}
		\State $\mathbf{U}\gets$ apply the $m$-th NN encoder to the $m$-dimension of $\mathbf{U}$ to map length-$k_m$ vectors to length-$n_m$ vectors
		\EndFor
		\State $\mathbf{C}_1\gets$ reshape each array in the batch $\mathbf{U}$ to a sequence of length $n=n_1n_2\cdots n_M$
		\State $\mathbf{C}\gets\texttt{PowerNormalizer}(\mathbf{C}_1)$
		\State \textbf{return} $\mathbf{C}$
	\end{algorithmic}
\end{algorithm}

\subsubsection{Encoder Architecture and Training Schedule}\label{sec_enc}
As shown in Fig. \ref{fig2}, we replace the two encoders by two FCNNs, namely $\mathcal{E}_1(\mathbf{\Phi}_1)$ and $\mathcal{E}_2(\mathbf{\Phi}_2)$, each parametrized with a set of weights $\mathbf{\Phi}_1$ and $\mathbf{\Phi}_2$, respectively. Each $j$-th encoder, $j=1,2$, has a input size of $k_j$, output size of $n_j$, and $L_{{\rm enc}j}$ hidden layers of size $N_{{\rm enc}j}$. Upon receiving a batch of information sequences, each reshaped to ($k_2\times k_1$)-dimensional arrays, the first encoder maps each length-$k_1$ row to a length-$n_1$ real-valued vector, resulting in a tensor $\mathbf{U}^{(1)}_{B\times k_2 \times n_1}$. The second encoder NN then maps each real-valued length-$k_2$ vector in the second dimension of $\mathbf{U}^{(1)}$ to a length-$n_2$ real-valued vector. Note that the mappings here are, in general, nonlinear mappings, and the resulting code is a nonlinear and non-binary code. At the end, the power normalization, to be clarified in Section \ref{impl_detl}, will be applied to the codewords.

The encoder training schedule at each epoch is as follows. First, a batch of $B$ length-$k_1k_2$ binary information sequences will be reshaped to a tensor $\mathbf{U}_{B\times k_2 \times k_1}$ to fit the encoding NNs. After encoding the input, the real-valued codewords will be passed through the AWGN channel and then will be decoded using the NN decoder (explained in Section \ref{sec_dec}), that is fixed throughout the encoder training schedule, to get the batch of decoded codewords $\hat{\mathbf{U}}_{B\times k_2k_1}$ (after appropriate reshaping). By computing the loss $\mathcal{L}(\mathbf{U},\hat{\mathbf{U}})$ between the transmitted and decoded sequences  and backpropagating the loss to compute its gradients, the encoder optimizer takes a step to update the weights of the NN encoders $\mathcal{E}_1(\mathbf{\Phi}_1)$ and $\mathcal{E}_2(\mathbf{\Phi}_2)$. This procedure will be repeated $T_{\rm enc}$ times while each time only updating the encoder weights for a fixed decoder model. 

The ProductAE encoder architecture and training schedule, for general $M$-dimensional ProductAEs, are summarized in Algorithms \ref{alg_PAE_training} and \ref{alg_PAE_Enc}.
 The function $\texttt{PowerNormalizer}$ in Algorithm \ref{alg_PAE_Enc} refers to the power normalization function.
More specifically, in order to ensure that the average power per coded symbol is equal to one and thus the average SNR is $1/\sigma^2$, the length-$n$ real-valued encoded sequences $\mathbf{c}=(c_1,c_2,…,c_n)$ are normalized as 
\begin{align}\label{norm}
	\mathbf{c}'=\sqrt{n}\mathbf{c}/||\mathbf{c}||_2.
\end{align}

\subsubsection{Decoder Architecture and  Training Schedule}\label{sec_dec}

\begin{figure*}[t]
	\centering
	\includegraphics[trim=2cm 10.5cm 2cm 10.8cm,width=6.8in]{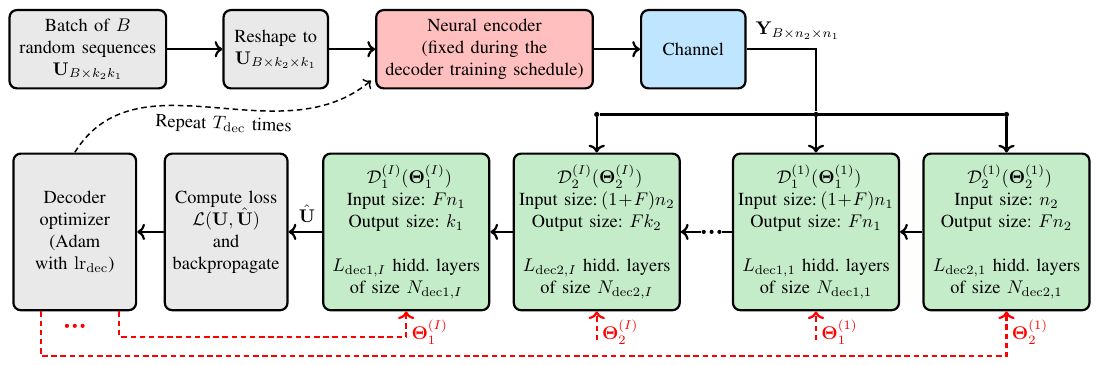}
	\vspace{-0.5cm}
	\caption{Decoder training schedule. The architecture of the 2D ProductAE decoder is also depicted. $I$ iterations of the decoder are replaced by $2I$ neural decoders $\mathcal{D}^{(i)}_j(\mathbf{\Theta}^{(i)}_j)$, $i=1,...,I$, $j=1,2$, each parametrized by the set of trainable weights $\mathbf{\Theta}^{(i)}_j$.}
	\label{fig3}
		\vspace{-0.1cm}
\end{figure*}

As shown in Fig. \ref{fig3}, we replace the pair of decoders at each $i$-th iteration, $i=1,\cdots I$, by a pair of distinct FCNNs resulting in $2I$ decoding NNs $\mathcal{D}^{(i)}_j$, $j=1,2$, each parametrized by a set of weights $\mathbf{\Theta}^{(i)}_j$. The architecture of each FCNN decoder is depicted in Fig. \ref{fig3}. Each NN decoder $\mathcal{D}^{(i)}_j$ has $L_{{\rm dec}j,i}$ hidden layers of size $N_{{\rm dec}j,i}$. Generally speaking, the first $I-1$ pairs of NN decoders work on input and output sizes of the length of coded bits $n_j$, while the last pair reverts the encoding operation by reducing the lengths from $n_j$ to $k_j$. Also, as further clarified in Section \ref{modif}, some of the decoders may take multiple length-$n_j$ vectors as input, and output multiple copies of length-$n_j$ vectors. 

During the decoder training schedule, at each epoch, first a batch of $B$ reshaped binary information sequences will be encoded using the fixed NN encoder. The real-valued codewords will be passed through the AWGN channel, with a carefully selected (range of) decoder training SNR, to generate the noisy observations of codewords.The batch of decoded information sequences  $\hat{\mathbf{U}}_{B\times k_2k_1}$ then is obtained by passing the noisy observations, i.e., the channel output, through the NN decoder.
 After computing the loss and its gradients, the decoder optimizer then updates the weights of the NN decoders. This procedure is repeated $T_{\rm dec}$ times while each time only updating the decoder weights for a fixed encoder model. The decoder training schedule is summarized in Algorithm \ref{alg_PAE_training}.

\subsection{Additional Modifications}\label{modif}
In this section, we briefly present several additional modifications to the ProductAE architecture which are essential to achieving excellent training performances.
Besides the modifications presented here, we also explored several other design and training strategies, some of which significantly helped in 
improving the training performance, including fine-tuning with very large batch sizes and considering a range for the decoder training SNR, to be discussed in Section \ref{impl_detl}. Additional investigations on the training and design of ProductAEs are also provided in Appendix \ref{app1}.

\begin{algorithm}[t]
	\caption{Decoder Function of 2-Dimensional ProductAEs  $\texttt{ProductAE}\_\texttt{Dec}\_\texttt{2D}$}\label{alg_2DPAE_Dec}
	\textbf{Input:} Batch of noisy codewords $\mathbf{Y}$, 
	set of code dimensions $k_1,k_2$ and blocklengths $n_1,n_2$, 
	number of decoding iterations $I$, and decoder NNs\\
	\textbf{Output:} Batch of decoded sequences $\hat{\mathbf{U}}$.
	\vspace*{0.05in}
	\begin{algorithmic}[1]
		\State $\mathbf{Y}\gets \mathbf{Y}.\texttt{reshape}(B,n_1,n_2)$ \Comment reshape $\mathbf{Y}$ to a 3D tensor
		\If{$I==1$}
		\State $\mathbf{Y}_{2,2}^{\rm in}\gets \mathbf{Y}$
		\Else
		\For{$i=1,\cdots, I-1$}
		
		\If{$i==1$}
		\State $\mathbf{Y}_2\gets \mathcal{D}_2^{(1)}(\mathbf{Y}).\texttt{reshape}(B,Fn_1,n_2)$ 
		\Else
        \State $\mathbf{Y}^{\rm out}_{2}\gets \mathcal{D}_2^{(i)}(\mathbf{Y}_{2,2}^{\rm in})$
        \State $\mathbf{Y}_{2}\gets (\mathbf{Y}^{\rm out}_{2}-\mathbf{Y}_{2,1}^{\rm in}).\texttt{reshape}(B,Fn_1,n_2)$
		\EndIf
		
		\State $\!\mathbf{Y}_1^{\rm in}\!\gets\! \texttt{concatenate}([\mathbf{Y}\!,\!\mathbf{Y}_2],\!1).\texttt{permute}(0,\!2,\!1)\!$
		\State $\mathbf{Y}_1\gets \mathcal{D}_1^{(i)}(\mathbf{Y}_1^{\rm in}).\texttt{permute}(0,2,1)$
		\State $\mathbf{Y}_{2,1}^{\rm in}\gets (\mathbf{Y}_1-\mathbf{Y}_2).\texttt{reshape}(B,n_1,Fn_2)$
		\State $\mathbf{Y}_{2,2}^{\rm in}\gets\texttt{concatenate}([\mathbf{Y},\mathbf{Y}_{2,1}^{\rm in}],2)$
		\EndFor
		\EndIf
		
		\State $\mathbf{Y}_{2}\gets \mathcal{D}_2^{(I)}(\mathbf{Y}_{2,2}^{\rm in}).\texttt{reshape}(B,Fn_1,k_2)$
		\State $\mathbf{Y}_1\gets \mathcal{D}_1^{(I)}(\mathbf{Y}_2.\texttt{permute}(0,2,1))$
		\State $\hat{\mathbf{U}}\gets \mathbf{Y}_1.\texttt{reshape}(B,k_1k_2)$
		\State \textbf{return} $\hat{\mathbf{U}}$
	\end{algorithmic}
\end{algorithm}

\subsubsection{Heterogeneous ProductAEs}\label{hetProductAE} Given that, at each decoding iteration, the decoding is first performed by $\mathcal{D}_2$ and then by $\mathcal{D}_1$, the decoder $\mathcal{D}_2$ is expected to observe noisier (less reliable) codewords than $\mathcal{D}_1$. Therefore, it is logical to use more powerful codes for $\mathcal{C}_2$ compared to $\mathcal{C}_1$. This can be done, e.g., by considering larger blocklengths or lower rates for $\mathcal{C}_2$ compared to $\mathcal{C}_1$. Our extensive training experiments (some of which presented in Section \ref{experiments}) confirm significant gains for such heterogeneous designs of ProductAEs.

\subsubsection{Channel Output Injection}\label{chnl_inj}
Following the structure of the SISO decoder for classical product codes \cite{pyndiah1998near}, in addition to the output from the previous decoder, we also pass the channel output to all NN decoders. In this case, as shown in Fig. \ref{fig3}, the first decoder $\mathcal{D}_2^{(1)}$ only takes the channel output as the input.
All the next $2I-2$ decoders take the channel output besides the output of the previous decoder, while the last decoder $\mathcal{D}_1^{(I)}$ only takes the output of the previous decoder as input (due to some issues with the size of the tensors being concatenated at the input of this decoder).
Our initial results show that this method \textit{alone} is not able to make a considerable improvement in the training.
However, the combination of this method with the next two modifications is experimentally proved to be very helpful.

\subsubsection{Adding Feature Size}\label{feature_size}
As shown in Fig. \ref{fig3}, each NN decoder, except the last one, outputs $F$ vectors as the soft information, where $F$ denotes the feature size. This is equivalent to asking each NN decoder to output $F$ different  estimates of the soft information instead of a single estimation. This is done by increasing the output size of the FCNNs. These estimates will then be carefully reshaped and processed before feeding the next decoder such that $F$ properly formed vectors are given to the next decoder (increasing the input size of FCNNs). Algorithm \ref{alg_2DPAE_Dec} summarizes this procedure in details.

\subsubsection{Subtraction of Soft Information}\label{subtrct}
We input the difference of the soft input from the soft output of each given decoder to the next decoder, i.e., the increments of the soft information is given to each next decoder. To this end,
the first decoder $\mathcal{D}_2^{(1)}$ only takes the channel output. Also, the second decoder directly receives the output of the first decoder since there was no soft input to $\mathcal{D}_2^{(1)}$. All the next decoders, except the last one (due to the dimension mismatch of arrays being subtracted),
 take the increments of soft information. Note that  this is, in essence, the principle of extrinsic information exchange between iterative decoders (similar to Turbo decoder).

We carried out extensive experiments to investigate the impacts of the aforementioned modifications. Our training results demonstrate that the combination of all last three methods together significantly improves the training performance.

The decoding algorithm for 2D ProductAEs in summarized in Algorithm \ref{alg_2DPAE_Dec}. Given that this paper builds upon 2D ProductAEs, for the sake of simplicity and clarity, we do not present the decoding algorithm for general $M$-dimensional ProductAEs. However, Algorithm \ref{alg_2DPAE_Dec} can be easily generalized to higher-dimensional ProductAEs.
In Algorithm \ref{alg_2DPAE_Dec}, the function $\texttt{reshape}$ reshapes the tensor upon which the function is being applied to the shape specified by the function arguments. Moreover, $\texttt{concatenate}([\mathbf{X},\mathbf{Y}],l)$ constructs a larger tensor by concatenating two tensors $\mathbf{X}$ and $\mathbf{Y}$ in dimension $l$. Finally, $\texttt{permute}$ returns a new tensor by permuting the dimensions of the original tensor according to the specified arguments.

\section{Experiments}\label{experiments}
In this section, we first briefly discuss some additional implementation aspects, and then provide the results of  training experiments over AWGN channels. The training results over non-AWGN channels will be presented in Section \ref{nonAWGN}.

\subsection{Additional Implementation Details}\label{impl_detl}

\subsubsection{Training with Very Large Batch Sizes}\label{subsec_B}
It is well known that increasing the batch size can significantly improve the training performance by averaging out the random noise \cite{jiang2019turbo}. Throughout our experiments, we use relatively large batches of size at least $B=5000$. 
Additionally, our experiments show that it is very beneficial to fine-tune the models with very large batch sizes at the end. To this end, throughout our experiments, once the training loss, for a given batch size, saturates with respect to the number of epochs, we load the best model and run a few number of epochs with a significantly large batch size to make sure that the training loss is saturated with respect to the batch size (i.e., it does not drop anymore by increasing the batch size). 

In order to handle training with very large batch sizes while avoiding memory shortage, we first run $L$ smaller batches of size $B_s$ while accumulating the gradients of loss without updating the optimizer across these batches. Note that
\begin{align}\label{loss}
\mathcal{L}(\mathbf{U},\hat{\mathbf{U}})&=\frac{1}{LB_s}\sum_{i=1}^{LB_s}l(u,\hat{u}_i)\nonumber\\
&=\frac{1}{L}\sum_{l=1}^{L}\frac{1}{B_s}\sum_{i_l=1}^{B_s}l(u,\hat{u}_{i_l})\nonumber\\
&=\frac{1}{L}\sum_{l=1}^{L}\mathcal{L}_l(\mathbf{U}_l,\hat{\mathbf{U}_l}),
\end{align}
where $\mathcal{L}_l(\mathbf{U}_l,\hat{\mathbf{U}_l})$ refers to the loss at each $l$-th smaller batch. Therefore, accumulating the loss per smaller batch, normalized to the number $L$ of smaller batches, is equivalent to having a single large batch of size $LB_s$.

\subsubsection{Optimizer, Activation Function, and Loss Function}\label{subsec_opt}
The following are used as the optimizer, activation function, and loss function throughout our experiments.
\begin{itemize}
	\item \textit{Optimizer:} We use Adam optimizer with learning rate ${\rm lr_{enc}}$ for the encoder and ${\rm lr_{dec}}$ for the decoder.
	\item \textit{Activation Function:} SELU (scaled exponential linear units) is applied as the activation function to all hidden layers. No activation function is applied to the output layer of any encoder and decoder NNs.
	\item \textit{Loss Function:} \texttt{BCEWithLogitsLoss()} is adopted as the loss function, which combines a ``sigmoid layer'' and the binary cross-entropy (BCE) loss in one single class in a more numerically stable fashion \cite{nn_loss}. More specifically, the loss function $\mathcal{L}(\mathbf{u}, \hat{\mathbf{u}})$ between two length-$k$ vectors $\mathbf{u}:=(u_i)_{1\times k}$ and $\hat{\mathbf{u}}:=(\hat{u}_i)_{1\times k}$ is defined as 
	\begin{align}
		\mathcal{L}(\mathbf{u}, \hat{\mathbf{u}})=-\frac{1}{k}\sum_{i=1}^k\big[&u_i\log\sigma(\hat{u}_i)\nonumber\\
		&+(1-u_i)\log\left(1-\sigma(\hat{u}_i)\right)\big],
	\end{align}
where $\sigma(x):=1/(1+{\rm e}^{-x})$ refers to the sigmoid function.
\end{itemize}

\subsubsection{Encoder and Decoder Training Iterations and SNRs}\label{subsec_snr} The encoder and decoder training schedules were discussed in Section \ref{proposed}. Throughout our experiments, we use $T_{\rm enc}=100$ and $T_{\rm dec}=500$. Also, at each encoder or decoder training iteration, the training SNR (equivalently, the noise variance of the channel) is carefully chosen. In particular, unless explicitly mentioned, we use a single-point of $\gamma$ \si{dB} for the encoder training SNR but a range of $[\gamma-2.5,\gamma+1]$ \si{dB} for the decoder training SNR. Therefore,
during each encoder training iteration, a single SNR is used to generate all noise vectors of the batch. However, during each iteration of the decoder training schedule, $B$ random values are picked uniformly from the interval $[\gamma-2.5,\gamma+1]$ \si{dB} to generate $B$ noise vectors of appropriate variances.


\subsubsection{Hyper-Parameter Values} Unless explicitly mentioned, the following values will be used to train the ProductAE models throughout the paper. We use $B=5000$, ${\rm lr_{enc}}={\rm lr_{dec}}=2\times 10^{-4}$, $I=4$,\footnote{While the experiments throughout the paper are obtained by choosing $I=4$ decoding iterations, as will be shown in Fig. \ref{impact_I}, some additional improvements are possible by choosing larger values of $I$.} and $F=3$. All FCNNs have $7$ hidden layers except the last pair of decoders which have $9$ hidden layers. Also, the size of all hidden layers is set to $N_{\rm enc}=200$ for the encoder FCNNs and $N_{\rm dec}=250$ for the decoder FCNNs.

\subsection{Training Results}
In this section, we present the training experiments for ProductAEs of different dimensions and lengths over AWGN channels. We show consistent gains for the performance of ProductAEs of dimensions $k=100$, $200$, and $300$ bits (compared to the benchmark performance of polar codes of same dimensions and lengths), thus demonstrating the potentials of ProductAE architecture in training large autoencoders. We also show meaningful gains for ProductAE compared to state-of-the-art classical and neural codes. 

Throughout the paper, we use the notation $(n,k)^{\otimes 2}$ and $(n_1,k_1)\otimes(n_2\times k_2)$ to denote 2D ProductAEs having component codes of identical and non-identical parameters, respectively.
Moreover, to construct a polar code of an arbitrary length $n$, we use random puncturing\footnote{Other rate matching approaches, such as shortening and repetition \cite{bae2019overview} are also possible. However, here, we are mostly interested in the relative performance of different ProductAEs.
Later in Fig. \ref{fig_classical}, we provide a thorough comparison with classical codes.}
 (except in Fig. \ref{fig_classical}, where the performance of the polar code is directly taken from \cite{jiang2019turbo}). More specifically, we first construct a longer polar code of length $2^{m_0}$, 
where $m_0=\ceil{\log_2 n}$, and then puncture the $2^{m_0}-n$ extra coded bits at random (this can be done by setting the log-likelihood ratios (LLRs) of the corresponding bit indices to zero). Additionally, to construct the polar code of any dimension $k$, the  approach in \cite{mori2009performance} is used to pick the $k$ bit-channels with the smallest BERs.


\subsubsection{ProductAE $(15,10)^{\otimes 2}$}\label{10_15}
\begin{figure}[t]
	\centering
	\includegraphics[trim=0.3cm 0.1cm 0 0,width=3.6in]{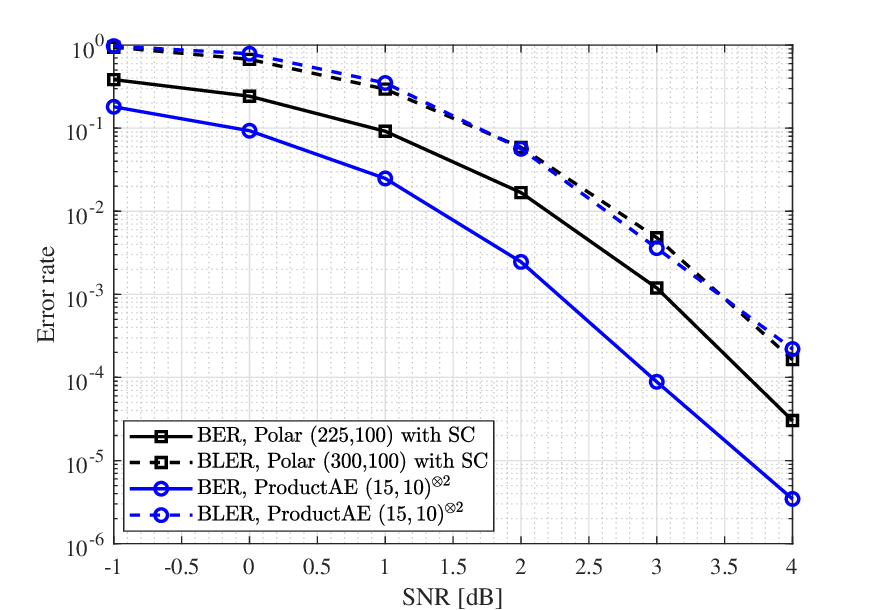}
	\caption{Testing result of ProductAE $(15,10)^{\otimes 2}$ compared to polar code $(225,100)$ under SC decoding. $\gamma=3$ dB is used for the encoder training SNR.}
	\label{fig4}
	\vspace{-0.25cm}
\end{figure}

Fig. \ref{fig4} presents the performance of ProductAE $(15,10)^{\otimes 2}$, where a single model is tested across all ranges of SNRs. As a benchmark, the ProductAE performance is compared to a polar code of the same length, i.e., with parameters $(225,100)$, under SC decoding.
 As seen, our trained ProductAE $(15,10)^{\otimes 2}$ beats the BER performance of the equivalent polar code with a significant margin over all ranges of SNR. Also, even though the loss function is a surrogate of the BER, our ProductAE is able to achieve almost the same BLER as the polar code.

\subsubsection{ProductAE $(21,14)^{\otimes 2}$}\label{14_21}
\begin{figure}[t]
	\centering
	\includegraphics[trim=0.3cm 0.1cm 0 0,width=3.6in]{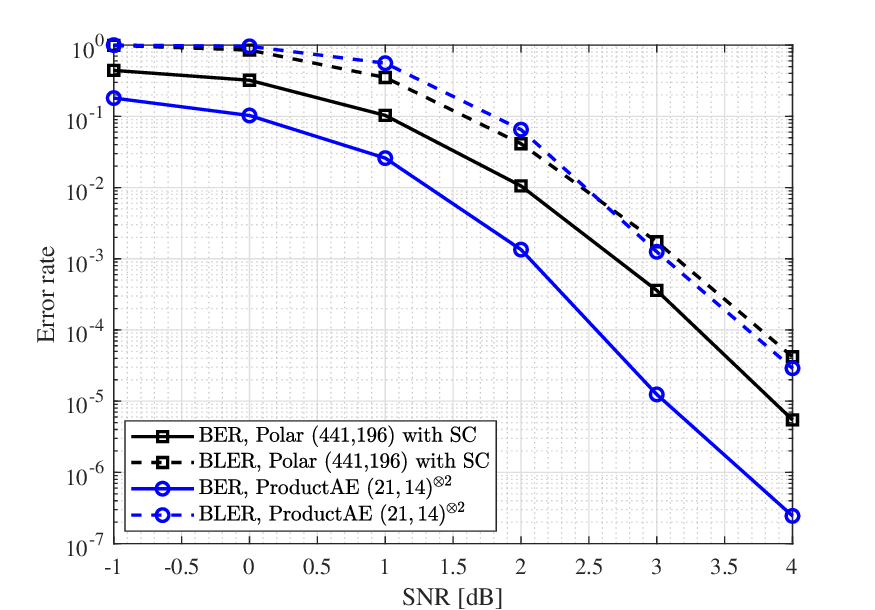}
	\caption{Testing result of ProductAE $(21,14)^{\otimes 2}$ compared to polar code $(441,196)$ under SC decoding. $\gamma=3$ dB is used for the encoder training SNR.}
	\label{fig5}
	\vspace{-0.25cm}
\end{figure}

Fig. \ref{fig5} compares the performance of the trained ProductAE $(21,14)^{\otimes 2}$ with a polar code of parameters $(441,196)$
under SC decoding. 
It is observed
 that the moderate-length ProductAE $(21,14)^{\otimes 2}$ beats the BER performance of the equivalent polar code with a large margin over all ranges of SNR while also maintaining almost the same BLER performance.
The consistent gain observed after almost doubling the size of the code, compared to ProductAE $(15,10)^{\otimes 2}$, clearly demonstrates the ability of the proposed ProductAE architecture in enabling the training of large error-correction autoencoder.

\subsubsection{Impact of Code Heterogeneity}\label{het_results}
\begin{figure}[t]
	\centering
	\includegraphics[trim=0.3cm 0.1cm 0 0,width=3.6in]{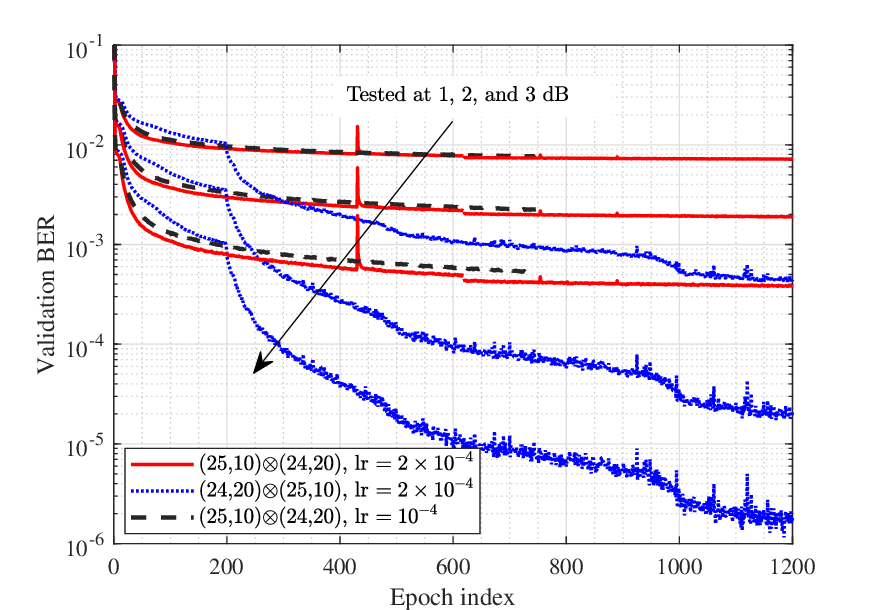}
	\caption{The impact of using a stronger code for $\mathcal{C}_2$ compared to $\mathcal{C}_1$ (heterogeneous ProductAE gain). $\gamma=1.25$ dB is used for all three training experiments.}
	\label{fig_het}
		\vspace{-0.25cm}
\end{figure}

As stated in Section \ref{hetProductAE}, it is beneficial to use stronger codes for $\mathcal{C}_2$ compared to $\mathcal{C}_1$, as the decoding is first performed via $\mathcal{D}_2$. To illustrate this, in Fig. \ref{fig_het} we are comparing the training profile of two ProductAEs: 1) ProductAE $(24,20)\otimes(25,10)$, which has almost twice smaller rate for $\mathcal{C}_2$ compared to $\mathcal{C}_1$; and 2) ProductAE $(25,10)\otimes(24,20)$, that is obtained by swapping the code parameters to give a stronger code for $\mathcal{C}_1$ compared to $\mathcal{C}_2$. It is observed that the former ProductAE design significantly outperforms the latter ProductAE, thus confirming our claim about heterogeneous ProductAE design. We performed additional training experiments (e.g., by changing the learning rate) on ProductAE $(25,10)\otimes(24,20)$   to confirm that the poor performance is not as a result of hyper-parameters but as a result of poor architecture. Our additional results in this section further highlight the advantages of heterogeneous designs of ProductAEs.

\subsubsection{ProductAE $(15,10)\otimes(20,10)$}\label{het_k100}
\begin{figure}[t]
	\centering
	\includegraphics[trim=0.3cm 0.1cm 0 0,width=3.6in]{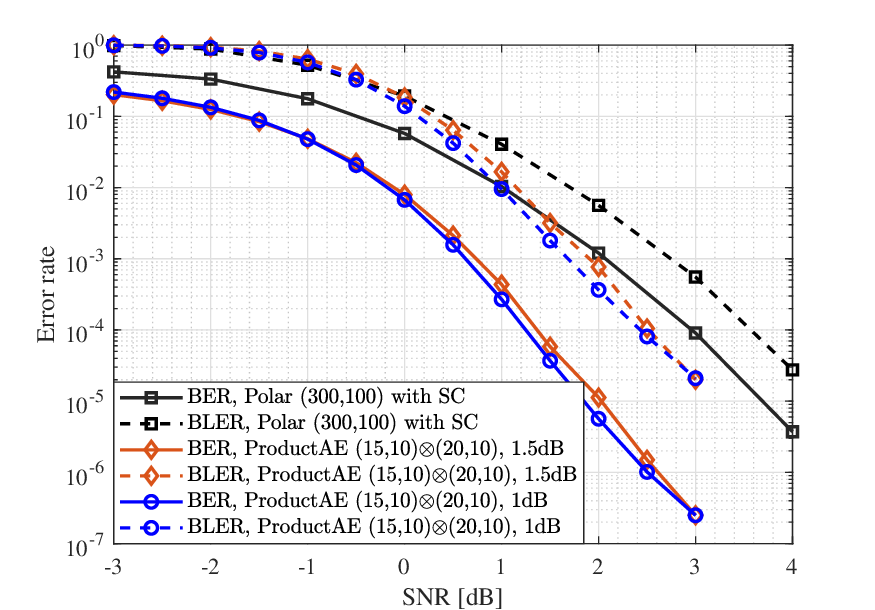}
	\caption{Testing result of ProductAE $(15,10)\otimes(20,10)$ compared to polar code $(300,100)$ under SC decoding. Two different models are tested for the ProductAE, one trained under $\gamma=1.5$ dB (the orange-diamond curves) and the other trained under $\gamma=1$ dB and fine-tuned with $\gamma=1.5$ dB (the blue-circle curves).}
	\label{hetro_k100}
		\vspace{-0.25cm}
\end{figure}

In Fig. \ref{hetro_k100}, the performance of the (heterogeneous) ProductAE $(15,10)\otimes(20,10)$ is compared to the performance of the equivalent polar code of parameters $(300,100)$. Given the heterogeneous design, significant gains, e.g., more than $1.5$ dB, are achieved compared to the polar code under SC decoding. Even more interestingly, this ProductAE is able to also deliver an excellent BLER performance, e.g., it achieves nearly $1$ dB gain compared to the polar code at the BLER of $10^{-4}$. 
Two different models are tested for the ProductAE in Fig. \ref{hetro_k100}. A slightly better performance at the moderate ranges of the SNR is observed when training the ProductAE under $\gamma=1$ dB and then fine-tuning with $\gamma=1.5$ dB (the blue-circle curves) compared to training under $\gamma=1.5$ dB (the orange-diamond curves). Hereafter, when we refer to the performance of the ProductAE $(15,10)\otimes(20,10)$, we mean the performance of the former model, i.e., the blue-circle curves.

\subsubsection{ProductAE $(24,20)\otimes(25,10)$}\label{het_k200}
\begin{figure}[t]
	\centering
	\includegraphics[trim=0.3cm 0.1cm 0 0,width=3.6in]{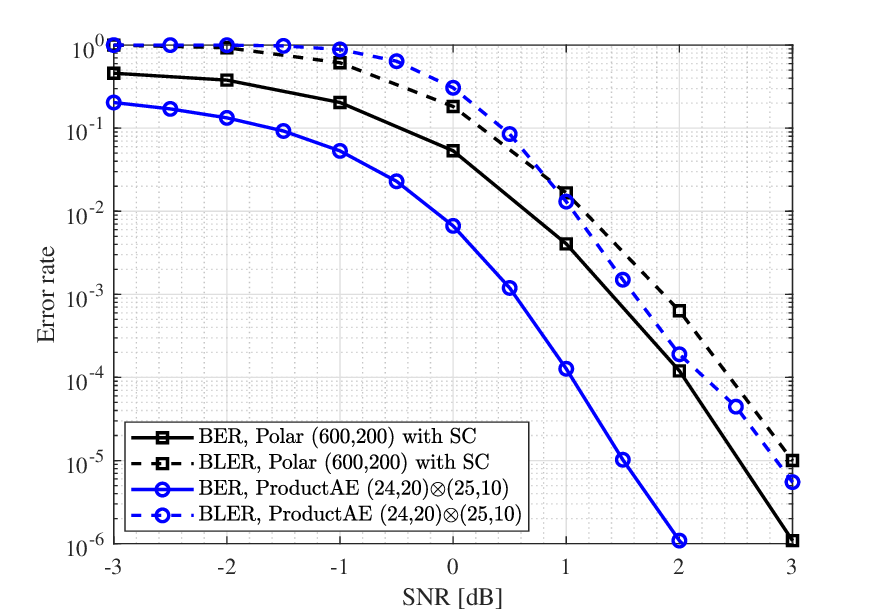}
	\caption{Testing result of ProductAE $(24,20)\otimes(25,10)$ compared to polar code $(600,200)$ under SC decoding. $\gamma=1.25$ dB is used for the encoder training SNR.}
	\label{hetro_k200}
		\vspace{-0.25cm}
\end{figure}

Fig. \ref{hetro_k200} compares the performance of the moderate-length ProductAE $(24,20)\otimes(25,10)$ with the equivalent polar code of parameters $(600,200)$. Despite doubling the size of the code compared to Fig. \ref{hetro_k100}, the ProductAE still achieves a significant (nearly $1$ dB) gain compared to the polar code. It is worth mentioning that we used the same network architecture (in terms of the number of layers and nodes) for both ProductAEs $(24,20)\otimes(25,10)$ and $(15,10)\otimes(20,10)$, which can be a reason for achieving a some higher gains for the latter ProductAE compared to polar codes. 

\subsubsection{ProductAE $(24,20)\otimes(25,15)$}\label{het_k300}
\begin{figure}[t]
	\centering
	\includegraphics[trim=0.3cm 0.1cm 0 0,width=3.6in]{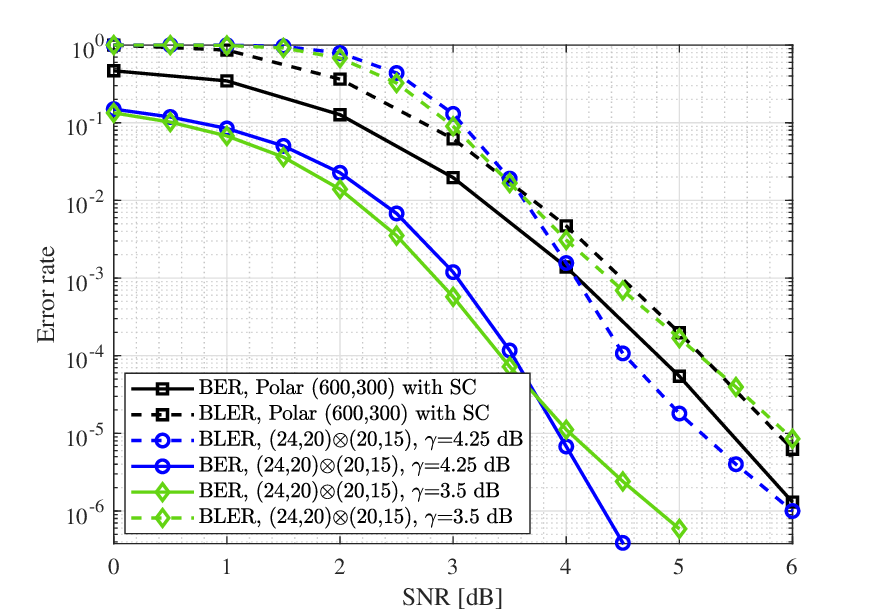}
	\caption{Testing result of ProductAE $(24,20)\otimes(25,15)$ compared to polar code $(600,300)$ under SC decoding. Two different encoder training SNRs of $\gamma=3.5$ dB and $\gamma=4.25$ dB are used for the ProductAE training.}
	\label{hetro_k300}
		\vspace{-0.25cm}
\end{figure}

Fig. \ref{hetro_k300} depicts the performance of the ProductAE $(24,20)\otimes(25,15)$ trained under two different encoder training SNRs of $\gamma=3.5$ dB and $\gamma=4.25$ dB. Despite having a relatively large code dimension of $k=300$, the ProductAE trained under $\gamma=4.25$ dB achieve significant gains compared to the equivalent polar code of parameters $(600,300)$. For example, nearly $1.8$, $1.5$, and $0.7$ dB gains are achieved compared to the polar code at the BERs of $10^{-6}$, $10^{-5}$, and BLER of $10^{-4}$, respectively. 
 To the best of our knowledge, this is the first-ever work to report successful training of pure neural (encoder, decoder) pairs, i.e., a completely new class of nonlinear neural codes paired with efficient decoders, over such relatively large code dimensions.
 
 To illustrate the impact of the training SNR, the performance of the ProductAE trained under $\gamma=3.5$ dB is also shown in Fig. \ref{hetro_k300}. While ProductAE demonstrates an excellent generalization to testing SNRs (all the results presented in the paper are with respect to testing of a single model across a wide range of SNR), still a varied performance may be achieved by changing the training SNR. For example,
 a slightly better performance at lower SNRs can be achieved by reducing the training SNR to $\gamma=3.5$ dB from $\gamma=4.25$ dB, however, at the expense of an inferior performance at higher SNR ranges. Therefore, the choice of training SNRs can depend on the application, e.g., whether the code is targeted for 
 low error rates (high reliability) or for moderate error rates. 
Making ProductAEs to further generalize across the SNR is a viable direction for future research. 

\subsubsection{Comparisons with Classical Codes}\label{classical_codes}
\begin{figure}[t]
	\centering
	\includegraphics[trim=0.3cm 0.1cm 0 0,width=3.6in]{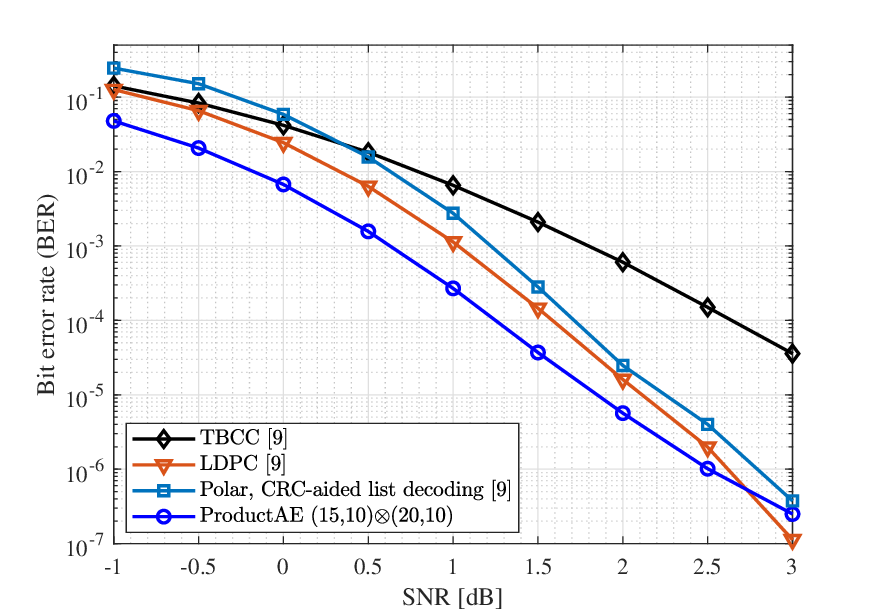}
	\caption{Performance comparison of ProductAE $(15,10)\otimes(20,10)$ with classical codes of equivalent parameters $n=300$ and $k=100$.}
	\label{fig_classical}
	\vspace{-0.25cm}
\end{figure}

Fig. \ref{fig_classical} compares the ProductAE's performance with state-of-the-art classical codes, namely polar code under cyclic redundancy check (CRC)-aided list decoding, LDPC, and tail-biting convolutional code (TBCC). All these three codes have parameters $(300,100)$, and their performance are directly extracted from \cite[Fig. 1]{jiang2019turbo}. 
Remarkably,  ProductAE $(15,10)\otimes (20,10)$ outperforms these classical codes with a significant margin over all ranges of SNR. For example, nearly $1.35$ dB, $0.45$ dB, and $0.35$ dB gains are achieved compared to TBCC, polar code, and LDPC, respectively, at the BER of $10^{-4}$. 
The results in this figure further highlight the potentials of ProductAEs in enabling neural codes of large dimensions that beat the state-of-the-art performance with applications in different domains, including the design of next generations of wireless networks.

\subsubsection{Comparisons with TurboAE}\label{turboAE}
\begin{figure}[t]
	\centering
	\includegraphics[trim=0.3cm 0.1cm 0 0,width=3.6in]{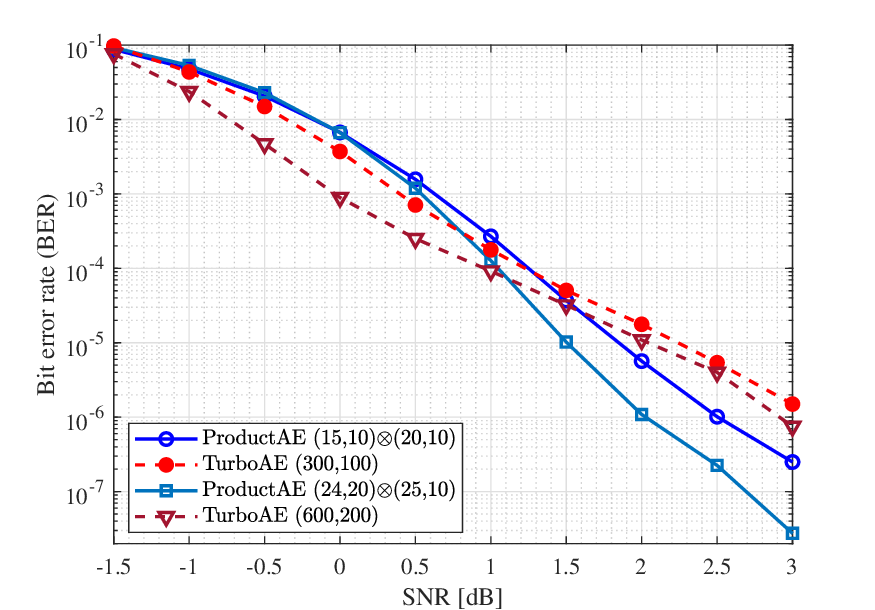}
	\caption{Performance comparison of rate-$1/3$ ProductAEs and TurboAEs of lengths $n=300$ and $600$.}
	\label{fig_pae_tae}
	\vspace{-0.25cm}
\end{figure}

Fig. \ref{fig_pae_tae} compares the performance of ProductAE with a state-of-the-art neural code, namely TurboAE \cite{jiang2019turbo}. Although TurboAE $(300,100)$ is a very well designed and trained channel autoencoder, our rate-$1/3$ ProductAE $(15,10)\otimes(20,10)$ of similar parameters is able to achieve a superior performance compared to TurboAE. In particular, ProductAE $(15,10)\otimes(20,10)$ (significantly) outperforms TurboAE $(300,100)$ for BER values smaller than $10^{-4}$ while having a slightly inferior performance at lower SNRs. Fig. \ref{fig_pae_tae} also provides the comparison of the two autoencoders for $k=200$. TurboAE $(600,200)$  result is obtained by testing a model trained for TurboAE $(300,100)$ (due to difficulty in directly training the TurboAE for large $k$'s). It is observed that TurboAE $(600,200)$ error curve has a similar slope to that of TurboAE $(300,100)$ and it only achieves a minimal improvement compared to the latter TurboAE. On the other hand, ProductAE $(24,20)\otimes(25,10)$ achieves a higher error rate slope compared to ProductAE $(15,10)\otimes(20,10)$ and delivers a much better performance, especially at higher SNRs, despite having a network architecture similar to the latter ProductAE, thus preserving the coding gain expected by doubling the blocklength. In comparison, our ProductAE $(24,20)\otimes(25,10)$ outperforms the performance of TurboAE $(600,200)$ by  $0.9$ \si{dB} and $0.5$ \si{dB} at the BERs of $10^{-6}$ and $10^{-5}$, respectively.
Besides the meaningful performance gains, ProductAE has several additional advantages compared to TurboAE, including the flexibility with the choice of code dimension $k$ and blocklength $n$, lower encoding and decoding complexity due to breaking down the encoding and decoding to smaller encoder and decoder components, highly parallel implementation (thus lower encoding and decoding latency), and the ability to train for larger code dimensions, among others.

\section{Robustness and Adaptivity Studies}\label{nonAWGN}
Among the major advantages of ML-driven classes of codes, compared to classical codes, are their robustness to changes in the environment as well as their ability to adapt to such changes. In this section, we study the robustness and adaptivity of ProductAEs to changes in the communication channel model by studying a sample ProductAE $(24,20)\otimes(25,10)$. 
\textit{Robustness}, here, refers to the ability of an autoencoder trained over a particular channel model to perform well on a different channel model without retraining. 
\textit{Adaptivity}, on the other hand, refers to the ability of an autoencoder to retrain for (i.e., adapt to) a different channel model with minimal retraining.

For the sake of brevity, we only consider a model that was originally trained over the AWGN channel, and then test its robustness and adaptivity to fast Rayleigh fading channels. 
In the case of fast Rayleigh fading, the received signal can be modeled as $\mathbf{y}=\boldsymbol{\alpha}\odot\mathbf{c}+\mathbf{n}$, where $\odot$ denotes the Hadamard (element-wise) product, and  $\boldsymbol{\alpha}$ is the length-$n$ vector of fading coefficients whose each $j$-th element follows a Rayleigh distribution as
\begin{align}
f_{A_j}(\alpha_j)=\frac{2\alpha_j}{\Omega_j}{\rm exp}(-\alpha_j^2/\Omega_j),
\end{align}
where $\Omega_j=\E[A_j^2]$. 
To make sure that fading neither amplifies nor attenuates the signal power (i.e., it only models the fluctuations), we set $\Omega_j=\E[A_j^2]=1$.
Note that the amplitude $A_j$ of a complex number $U_j+iV_j$, whose real and imaginary parts $U_j$ and $V_j$ are Gaussian random variables with mean zero and variance $\sigma_j^2$, has a Rayleigh distribution with parameter $\Omega_j=2\sigma_j^2$. Therefore, we can generate Rayleigh fading samples by picking independent samples $u_j$ and $v_j$ from a Gaussian 
distribution
with mean zero and variance $1/2$, and then obtaining $\alpha_j=\sqrt{u_j^2+v_j^2}$.

\begin{table*}[t]
	\centering
\caption{The description of the models used in the robustness and adaptivity studies for ProductAE $(24,20)\otimes(25,10)$.\label{table1}}
\begin{tabular}{M{1.5cm}||p{5.8in}}
	Model name & Model description \\ \hline\hline
	45003 & This model, which represents a base model for all robustness and adaptivity experiments in the paper, is obtained by first training over the AWGN channel with the batch size $B=6000$,  the encoder training SNR $\gamma=1.25$ dB, and the range $[-1.25,2.25]$ dB for the decoder training SNR,  then fine-tuning over the same channel with the same training SNRs but with the batch size $B=45000$, and finally picking the model that has the best validation BER at $3$ dB.\\ \hline
	AWGN-E6 & This model is obtained by loading the model 45003, and then fine-tuning that model with the encoder training SNR $\gamma=4$ dB, the range $[1,6]$ dB for the decoder training SNR, and batch size $B=45000$, over the AWGN channel, for $E=6$ epochs.\\ \hline
	AWGN-E11 & This model is obtained in a similar fashion to model AWGN-E6 but by picking the model after $E=11$ epochs.\\ \hline
	Rayleigh-E1 & This model is obtained by fine-tuning model 45003 over Rayleigh fading channels with $B=45000$, $\gamma=5$ dB for the encoder training SNR, the range $[2,7]$ dB for the decoder training SNR, and picking the model after only $E=1$ epoch.\\ \hline
	Full Retraining & This model (see Fig. \ref{fig_retrain}) is obtained by full retraining (from scratch) of a model over the Rayleigh fading channel with $B=6000$ (and $B=45000$ for fine-tuning), $\gamma=4$ dB for the encoder training SNR, and range $[2,6]$ dB for the decoder training SNR.\\\hline
\end{tabular}
\vspace{-0.2cm}
\end{table*}

\subsection{Robustness Studies}\label{rob}
\begin{figure}[t]
 	\centering
 	\includegraphics[trim=0.3cm 0.1cm 0 0,width=3.6in]{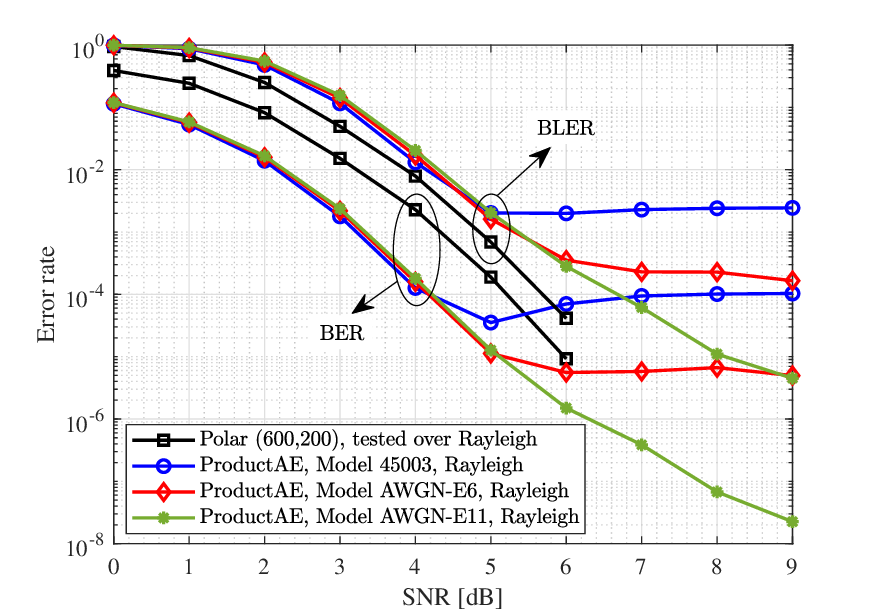}
 	\caption{Robustness of ProductAE: The testing performance over the Rayleigh fading channel is presented for three different models trained over the AWGN channel for the ProductAE $(24,20)\otimes(25,10)$ (see Table \ref{table1} for the description of the models).}
 	\label{fig_rob}
 		\vspace{-0.25cm}
 \end{figure}

Fig. \ref{fig_rob} shows the robustness of ProductAE by testing the performance over the Rayleigh fading channel for three different models, each trained over the AWGN channel for the ProductAE $(24,20)\otimes(25,10)$. The complete description of the models used throughout the robustness and adaptivity studies in the paper is provided in Table \ref{table1}. We first tested the performance of the Model 45003 (blue circle curves) that was trained over the AWGN channel specifically to achieve a good performance over such channels, e.g., by picking the encoder training SNR of $\gamma=1.25$ dB and decoder training SNR range of $[-1.25,2.25]$ dB. The testing results over the Rayleigh fading channel in Fig. \ref{fig_rob} show that this model achieves an excellent performance gain of more than $1$ dB compared to the polar code $(600,200)$ for the BER values larger than $10^{-4}$. However, the error rate curves saturate for larger SNR values, resulting in an error floor. We conjecture, based on our extensive training experiments 
 (see also Appendix \ref{app1_saturate} for some additional discussions) that the saturation of the performance is because of picking a model that is optimized for a given (range of) SNR but is not good for the subsequent larger values of SNR. In other words, we are picking the training SNRs of the encoder and decoder in a range that does not include any (significant number of) noisy samples with the large channel SNRs we are testing. As a result, the trained models completely overlook the larger values of SNR while only trying to optimize the models for the ranges of SNRs considered in training.
 
 \begin{figure}[t]
 	\centering
 	\includegraphics[trim=0.3cm 0.1cm 0 0,width=3.6in]{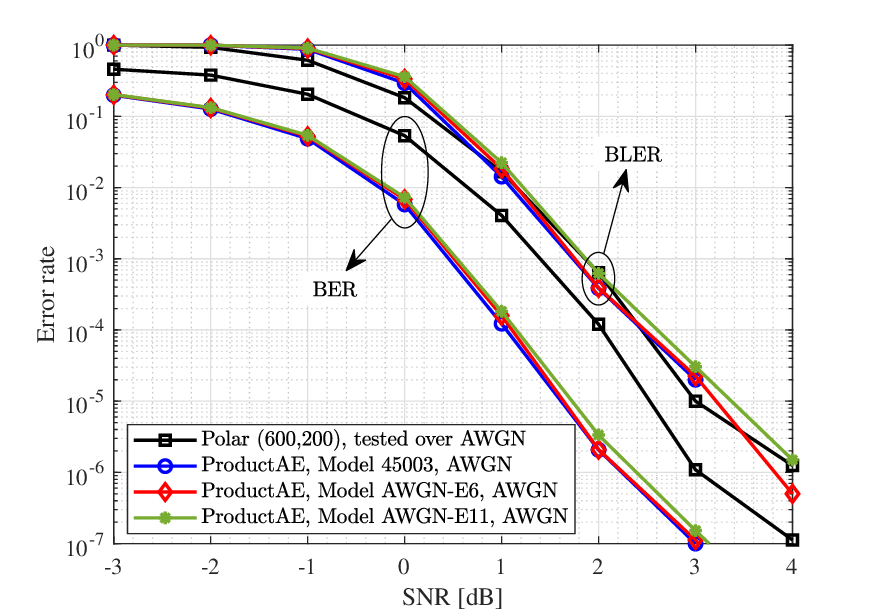}
 	\caption{Robustness of ProductAE: The performance of fine-tuned models, considered for the robustness experiments in Fig. \ref{fig_rob}, is shown over the AWGN channel to demonstrate that such fine-tunings do not degrade the performance over the AWGN channel.}
 	\label{fig_rob2}
 		\vspace{-0.25cm}
 \end{figure}
 
 Comparison of the results in Figs. \ref{hetro_k200} and \ref{fig_rob} illustrates that changing the transmission channel from AWGN to Rayleigh fading not only shifts the error rate curves to right (thus requiring higher values of SNR to achieve the same error rate) but also decreases the slope of the error rate curves. Given this observation and the above discussions, we propose fine-tuning Model 45003 with larger values (and wider ranges) of the training SNRs for a few number of epochs to alleviate the performance saturation behavior at higher SNRs. In Fig. \ref{fig_rob}, the testing results for fine-tuned models over the AWGN channel with the encoder training SNR $\gamma=4$ dB and the decoder training SNR range $[1,6]$ dB are presented after $E=6$ and $11$ epochs. It is observed that as we fine-tune Model 45003 for larger number of epochs, the saturation behavior decreases such that error floor almost vanishes after $E=11$ epochs. More importantly, Fig. \ref{fig_rob2} illustrates that such a fine-tuning does not degrade the performance of the fine-tuned models (compared to the base model 45003) over the AWGN channel as long as we do the fine-tuning for relatively small number of epochs. It is worth emphasizing that the fine-tuning is performed over the AWGN channel to obtain models that not only achieve an excellent performance over the AWGN channel (as confirmed in Fig. \ref{fig_rob2}) but also are robust when tested over Rayleigh fading channels. We believe, in general, such an approach is helpful in generating models that perform well over a set of channel dynamics.

\subsection{Adaptivity Studies}\label{adapt}
\begin{figure}[t]
	\centering
	\includegraphics[trim=0.3cm 0.1cm 0 0,width=3.6in]{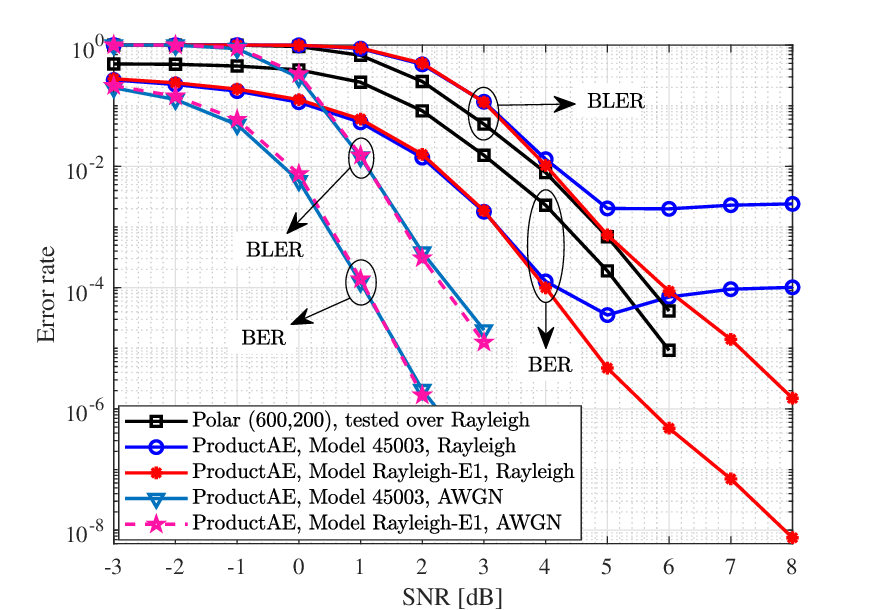}
	\caption{Adaptivity of ProductAE: ProductAE $(24,20)\otimes(25,10)$ is able to fully adapt to the Rayleigh fading channel by only one full epoch of fine-tuning over the fading channel (see Table \ref{table1} for the description of the models).}
	\label{fig_adapt}
		\vspace{-0.25cm}
\end{figure}

Fig. \ref{fig_adapt} shows the ability of ProductAE $(24,20)\otimes(25,10)$ to fully adapt to Rayleigh fading channels. In particular, only one full epoch of training of the model 45003 (that was trained over the AWGN channel) over the Rayleigh fading channel results in a model (named model Rayleigh-E1 in this paper) that achieves an excellent performance over the Rayleigh fading channel. Although the performance of the new model over the AWGN channel (i.e., the initial channel model) does not matter much under the adaptivity formulation, Fig. \ref{fig_adapt} shows that the model Rayleigh-E1 delivers as good performance as the model 45003 when tested over the AWGN channel. In other words, the new model after only one epoch of training over the new channel model not only achieves a great performance over that new channel  but  also can be used over the initial channel model without any loss on the performance. This suggests another potential approach in training models that perform well over multiple channel models. 

\begin{figure}[t]
	\centering
	\includegraphics[trim=0.3cm 0.1cm 0 0,width=3.6in]{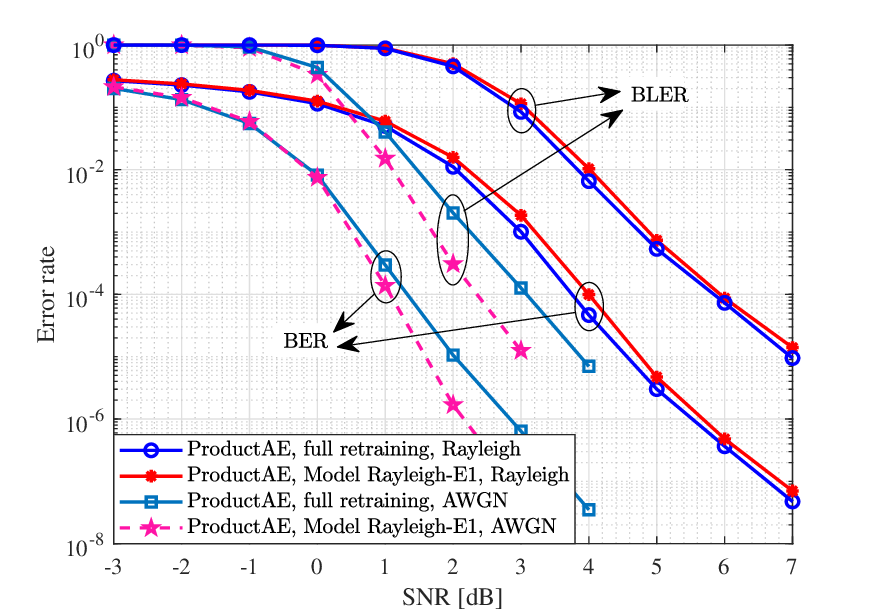}
	\caption{Performance of ProductAE $(24,20)\otimes(25,10)$ over the AWGN and Rayleigh fading channels after full retraining (from scratch) over the Rayleigh fading channel.}
	\label{fig_retrain}
		\vspace{-0.25cm}
\end{figure}

In order to further examine the performance of the model Rayleigh-E1, we fully re-trained (from scratch)  ProductAE $(24,20)\otimes(25,10)$ over the Rayleigh fading channel.
Fig. \ref{fig_retrain} shows that the model Rayleigh-E1
 delivers almost the same performance as this new fully re-trained model. This further demonstrates the adaptivity of ProductAEs. As expected, this new model achieves a worse performance, compared to the model Rayleigh-E1, over the AWGN channel. However, one might be able to follow similar approaches to adapt this model to the AWGN channel model.

\section{Conclusions}\label{conc}
In this paper, we presented the ProductAE architecture motivated by the goal of training large error-correction autoencoders based on smaller encoder and decoder components. 
This work is the first to explore training product codes, and is a pioneering work toward training large channel autoencoders.
We demonstrated significant performance gains, compared to state-of-the-art neural and classical codes, via training ProductAEs of dimensions as large as $k=300$. We also established the excellent robustness and adaptivity of ProductAEs to changes in the communication channel model.
It is worth mentioning that the proposed architecture and training procedures are general, and can be applied to train even larger ProductAEs than what was presented in this paper.
Given that ProductAE enables a low encoding and decoding complexity, flexible code rates and lengths, and low encoding and decoding latency  (thanks to a highly parallel implementation), among others, the achieved performance gains are truly remarkable, and call for extensive future research. This work paves the way toward designing larger neural codes that beat the state-of-the-art performance with potential applications in different domains, including the design of next generations of wireless networks.

\appendices
\section{Additional Investigations on the Design and Training of ProductAEs}\label{app1}
In this appendix, we briefly discuss
 our several additional 
investigations on the design and training of ProductAEs. 
Although our training experiments do not show much gains for some of the schemes presented in this appendix, we find the intuitions obtained from these studies helpful in understanding the design of ProductAEs as well as in defining guidelines for future research directions.

\begin{figure}[t]
	\centering
	\includegraphics[trim=0.3cm 0.3cm 0 0,width=3.6in]{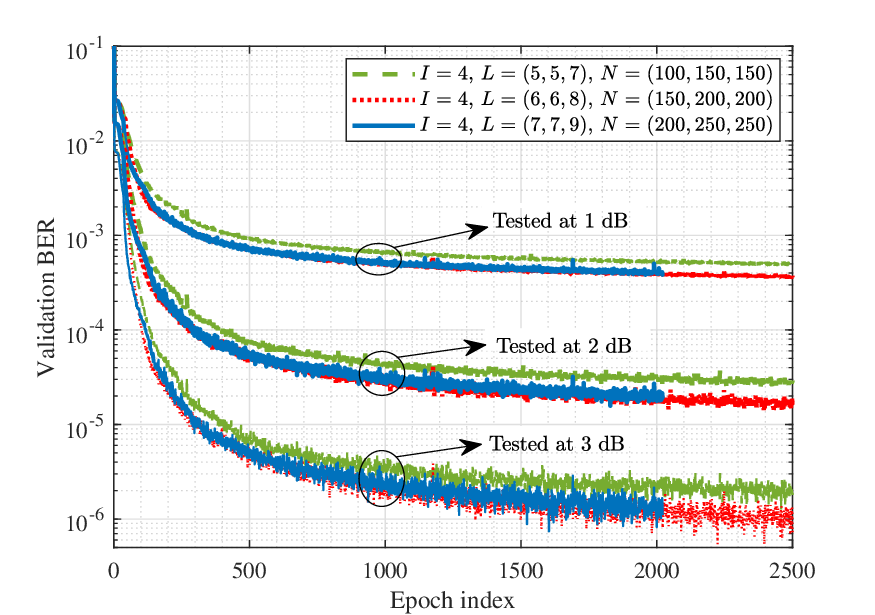}
	\caption{Impact of reducing the number of layers and nodes on the training performance of ProductAE $(15,10)\otimes(20,10)$. $\gamma_{\rm e}=1$ dB, $\gamma_{\rm d}\in[-1.5,2]$ dB, and ${\rm lr_{enc}}={\rm lr_{dec}}=3\times 10^{-4}$ are used for training.  The numbers in $L$ (and $N$) represent the number of hidden layers (and nodes) in the encoders, decoders (except the last pair), and the last pair of decoders, respectively.}
	\label{compRed}
		\vspace{-0.25cm}
\end{figure}
\subsection{Reduced Model Sizes}\label{complexity}
Throughout the paper, we mostly focused on optimizing the performance without paying much attention to the computational complexity and size of the network architectures. Two immediate approaches to reduce the size of the ProductAE encoding and decoding networks are 1) decreasing the number of hidden layers and nodes in the encoder and decoder FCNNs; and 2) using smaller number of decoding iterations $I$, that corresponds to the number of neural decoder pairs in the ProductAE architecture. In this section, we investigate the impact of these two approaches on the training of a sample ProductAE, namely ProductAE $(15,10)\otimes(20,10)$, trained under $\gamma_{\rm e}=1$ dB, $\gamma_{\rm d}\in[-1.5,2]$ dB, and ${\rm lr_{enc}}={\rm lr_{dec}}=3\times 10^{-4}$.

Fig. \ref{compRed} shows the impact of reducing the number of layers and nodes on the training performance of ProductAE. The three numbers in $L$ (and $N$) represent the number of hidden layers (and nodes) in the encoders, decoders (except the last pair), and the last pair of decoders, respectively. Recall that all training experiments in the paper were obtained by choosing $L=(7,7,9)$ and $N=(200,250,250)$ (the solid blue curves in Fig. \ref{compRed}). It is, however, observed that an identical training performance can be achieved by choosing simpler architectures with $L=(6,6,8)$ and $N=(150,200,200)$ (the dotted red curves). As summarized in Table \ref{table2}, the latter architecture has almost twice smaller number of learnable parameters compared to the former architecture. Moreover, only a minor performance degradation is observed by choosing $L=(5,5,7)$ and $N=(100,150,150)$ (the dashed green curves), which has nearly $4$ times smaller number of parameters compared to the benchmark architecture (i.e., the one with $L=(7,7,9)$ and $N=(200,250,250)$). Therefore, we argue that smaller number of layers and nodes, possibly together with model compression techniques in neural networks, can be used to reduce the complexity of our ProductAE architectures without (noticeably) degrading the performance.
\begin{table}[t]
	\centering
	\caption{Parameters of different architectures 
		considered in Section \ref{complexity}. 
		$|\mathbf{\Phi}_{\mathcal{E}}|$ and $|\mathbf{\Theta}_{\mathcal{D}}|$ denote the number of learnable parameters in the encoder and decoder, respectively.\label{table2}}
	\begin{tabular}{M{0.3cm}||M{1cm}||M{2cm}||M{1.1cm}||M{1.1cm}}
		$I$ & $L$ & $N$ & $|\mathbf{\Phi}_{\mathcal{E}}|$ & $|\mathbf{\Theta}_{\mathcal{D}}|$ \\
		\hline\hline
		
		$1$ & $(7,7,9)$ & $(200,250,250)$ & $493835$
		& $1030790$ \\\hline
		
		
		
		$2$ & $(7,7,9)$ & $(200,250,250)$ & $493835$
		& $1845645$ \\\hline			
		
		$3$ & $(7,7,9)$ & $(200,250,250)$ & $493835$ &
		$2660500$ \\\hline
		
		$4$ & $(5,5,7)$ & $(100,150,150)$  & $86535$
		& $942955$ \\\hline		
		
		$4$ &  $(6,6,8)$ & $(150,200,200)$ & $235085$
		& $1938755$ \\\hline
		
		$4$ & $(7,7,9)$ & $(200,250,250)$ & $493835$
		& $3475355$
		\\\hline		
		
		$5$ & $(7,7,9)$ & $(200,250,250)$ & $493835$ & $4290210$
		\\\hline
		
		$6$ & $(7,7,9)$ & $(200,250,250)$ & $493835$
		& $5105065$
		\\\hline
%
		
	\end{tabular}
\end{table}

\begin{figure}[t]
	\centering
	\includegraphics[trim=0.3cm 0.3cm 0 0,width=3.6in]{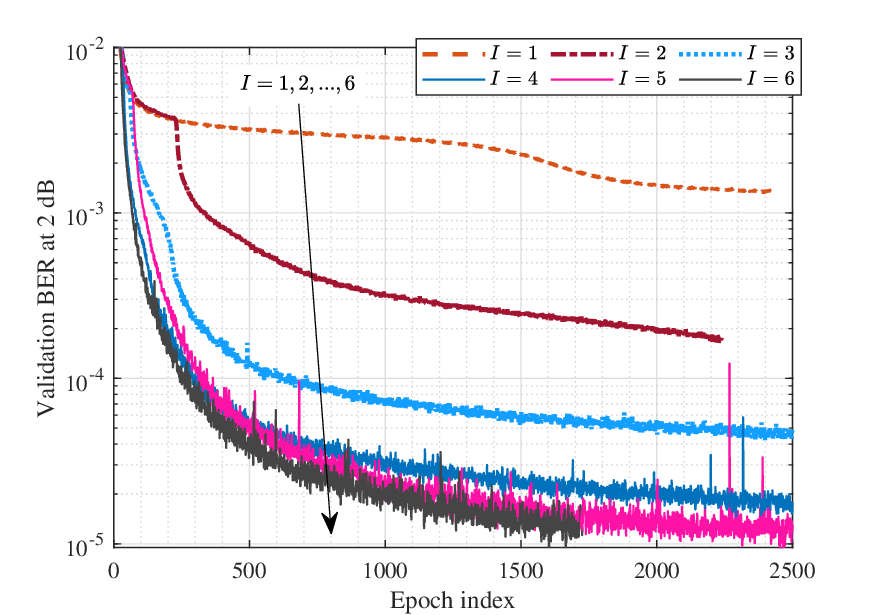}
	\caption{Impact of the number of decoding iterations $I$ (that corresponds to the number of neural decoder pairs in the ProductAE architecture) on the training of ProductAE $(15,10)\otimes(20,10)$. 
		$L=(7,7,9)$, $N=(200,250,250)$, $\gamma_{\rm e}=1$ dB, $\gamma_{\rm d}\in[-1.5,2]$ dB, and ${\rm lr_{enc}}={\rm lr_{dec}}=3\times 10^{-4}$ are used for training.}
	\label{impact_I}
		\vspace{-0.25cm}
\end{figure}

Fig. \ref{impact_I} illustrates the impact of decoding iterations $I$ on the training of ProductAE. Choosing smaller values of $I$ is observed to significantly degrade the training performance. Moreover, while we used $I=4$ throughout the paper, some improvements are still possible by choosing $I=5$ (and even $6$). Our additional training experiments demonstrate the saturation of the gain after $I=6$ iterations. Therefore, similar to iterative decoding of product codes \cite{pyndiah1998near}, a few decoding iterations are essential to guarantee a good training performance for ProductAEs.

\subsection{Multi-Schedule Training}
As shown in Fig. \ref{fig3},
the ProductAE decoding consists of $I$ decoding iterations, each 
represented 
 by a distinct pair of NNs. As such, the ProductAE decoder has a relatively complex network with much more learnable parameters compared to the ProductAE encoder. Therefore, one may wonder whether the decoder training 
can be improved
by performing several separate training schedules, similar to separating the encoder and decoder training schedules per epoch. To this end, we consider the following three approaches to investigate the multi-schedule training of the ProductAE decoder.
\begin{itemize}
	\item \textit{Scheme I:} In this scheme, different schedules are considered for each decoding iteration, leading to $I+1$ training schedules per epoch. At each schedule $i\in\{1,\cdots I\}$, only the decoders pair $\{\mathcal{D}^{(i)}_2,\mathcal{D}^{(i)}_1\}$ is updated/trained, each for $T_{\rm dec}^{(i)}$ iterations, while keeping all other encoding and decoding NNs fixed. Once all decoder NNs are updated, the encoder training schedule takes place to optimize the encoders pair $\{\mathcal{E}_1,\mathcal{E}_2\}$, for $T_{\rm enc}$ iterations, while keeping all decoder NNs fixed.
	
	\item \textit{Scheme II:} This scheme is a modification to Scheme I by adding a full decoder training schedule (i.e., training all decoder NNs together while keeping the encoder NNs fixed), consisting of $T_{\rm dec,s}$ iterations, at the beginning of the $I+1$ schedules in Scheme I. Therefore, each training epoch is composed of $I+2$ schedules.
	
	\item \textit{Scheme III:} This scheme is a modification to Scheme II by adding a full decoder training schedule, consisting of $T_{\rm dec,e}$ iterations, at the end of the
	$I+1$ decoder training schedules
	 in Scheme II. More specifically, in the first schedule, all decoders are trained together for $T_{\rm dec,s}$ iterations. During the next $I$ schedules, only a single pair of decoders, corresponding to a given decoding iteration, is updated. Afterward, another full decoder training schedule takes place for $T_{\rm dec,e}$ iterations. Finally, the encoders are updated for  $T_{\rm enc}$ iterations.
\end{itemize}

\begin{figure}[t]
	\centering
	\includegraphics[trim=0.3cm 0.3cm 0 0,width=3.6in]{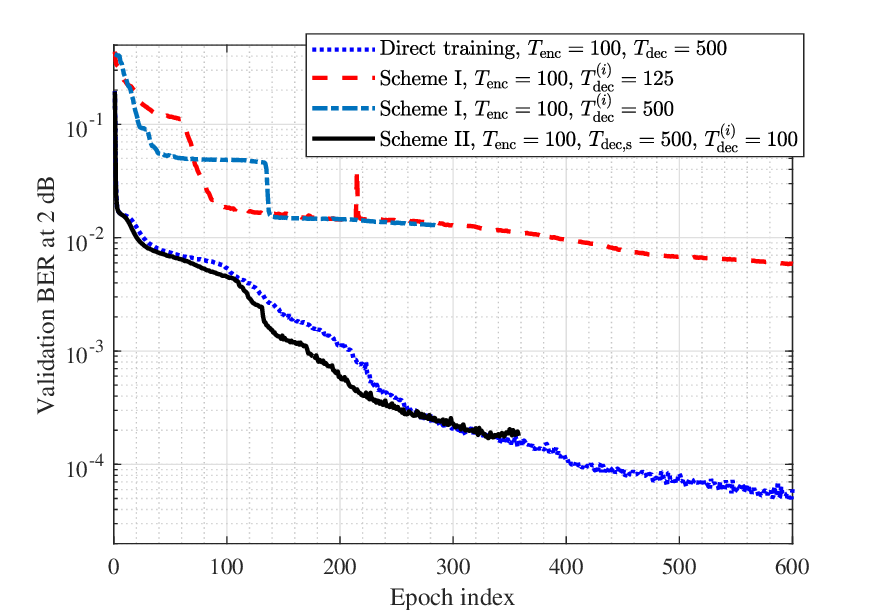}
	\caption{Multi-schedule training of the ProductAE $(24,20)\otimes(25,10)$ over the AWGN channel. Schemes I and II are compared to the direct training scheme, which refers to directly training the whole decoder (in a single schedule per epoch). The same set of hyper-parameters as Fig. \ref{hetro_k200} is used for the training.}
	\label{fig_multischedule1}
		\vspace{-0.25cm}
\end{figure}
Fig. \ref{fig_multischedule1} compares the multi-schedule training of the ProductAE $(24,20)\otimes(25,10)$ using Schemes I and II with the direct training scheme, which refers to directly training the whole decoder in a single schedule per epoch. 
It is observed that Scheme I significantly degrades the performance, compared to the direct training of the decoder, whether under the same total number of decoder training iterations (i.e., $T_{\rm dec}^{(i)}=125$) or under much larger number of decoder training iterations (i.e., $T_{\rm dec}^{(i)}=500$, which results in $4$ times larger number of decoder training iterations per epoch compared to direct training with $T_{\rm dec}=500$). In additional experiments, we also investigated fine-tuning pre-trained models 
 using Scheme I, and we did not observe improvements to pre-trained models.
On the other hand, Scheme II, with the help of a full decoder training schedule at the beginning, is observed to deliver a much better training compared to Scheme I. However, it is still unable to improve upon direct training, even under larger number of decoder training iterations.

\begin{figure}[t]
	\centering
	\includegraphics[trim=0.3cm 0.3cm 0 0,width=3.6in]{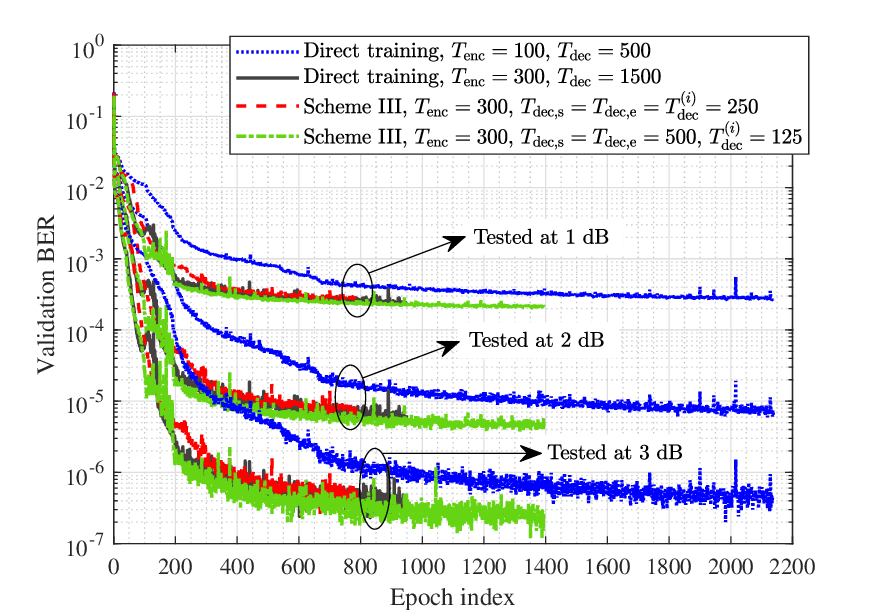}
	\caption{Multi-schedule training of the ProductAE $(24,20)\otimes(25,10)$. Scheme III is compared to the direct training scheme under different values of training iterations. The same set of hyper-parameters as Fig. \ref{hetro_k200} is used for the training, except that the learning rate is $4\times 10^{-4}$.}
	\label{fig_multischedule2}
		\vspace{-0.25cm}
\end{figure}
In Fig. \ref{fig_multischedule2}, Scheme III is compared to direct training of ProductAE $(24,20)\otimes(25,10)$ under different values of training iterations. Although Scheme III may initially seem very promising, additional training experiments in Fig. \ref{fig_multischedule2}, under the same total number of encoder and decoder training iterations, show that this is mainly due to the increased number of iterations by a factor of $3$. In other words, each training epoch of Scheme III in Fig. \ref{fig_multischedule2} consists of $3$ times larger number of iterations than the benchmark direct training under  $T_{\rm enc}=100$ and $T_{\rm dec}=500$. Therefore, it may take $3$ times smaller number of epochs to achieve the same training performance as the benchmark training scheme. Meanwhile, still some improvements are observed using Scheme III (and/or by increasing the number of iterations per epoch). Although all the results in the paper (except Figs. \ref{fig_multischedule1} and \ref{fig_multischedule2}) are without the use of multi-schedule training for the ProductAE decoder, we believe more in-depth research along the studies in this section can be helpful in further improving the training performance of ProductAEs.

\subsection{Error Floor at High SNRs}\label{app1_saturate}
As discussed in Section \ref{rob}, ProductAEs may experience error floor at high SNRs if the training SNRs are not chosen carefully. We believe the same thing can happen for any other classes of channel autoencoders. In this section, we explore this behavior in more details, and provide some heuristic approaches to address the issue.

\begin{figure}[t]
	\centering
	\includegraphics[trim=0.3cm 0.1cm 0 0,width=3.6in]{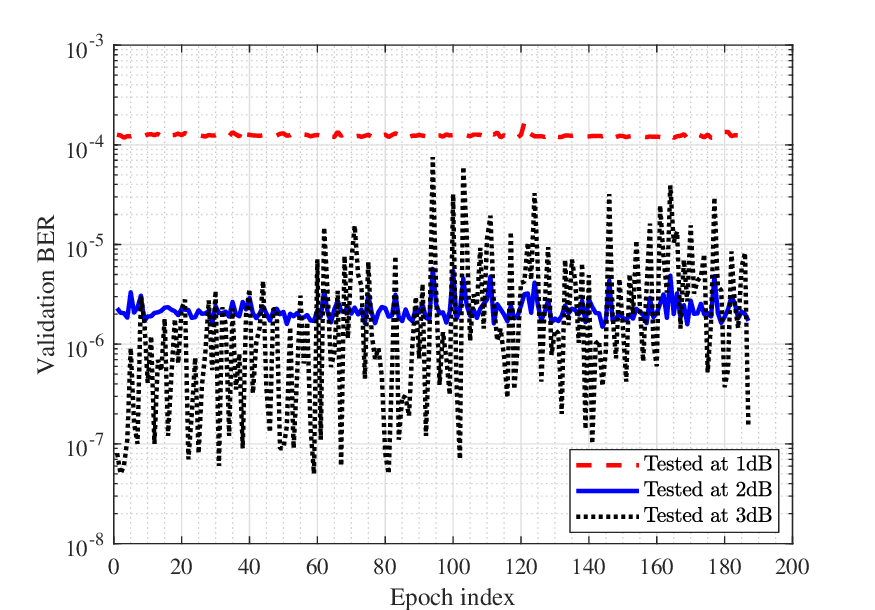}
	\caption{Fine-tuning of ProductAE $(24,20)\otimes(25,10)$ under $\gamma=1.25$ dB for the encoder training SNR (and range $[\gamma-2.5,\gamma+1]$ dB for the decoder training SNR). The validation BER at $3$ dB significantly fluctuates due to the lack of training samples from high SNRs. $5\times 10^{5}$ codewords are used for the validation of the BER across epochs.}
	\label{efloor1}
		\vspace{-0.25cm}
\end{figure}

\begin{figure}[t]
	\centering
	\includegraphics[trim=0.3cm 0.1cm 0 0,width=3.6in]{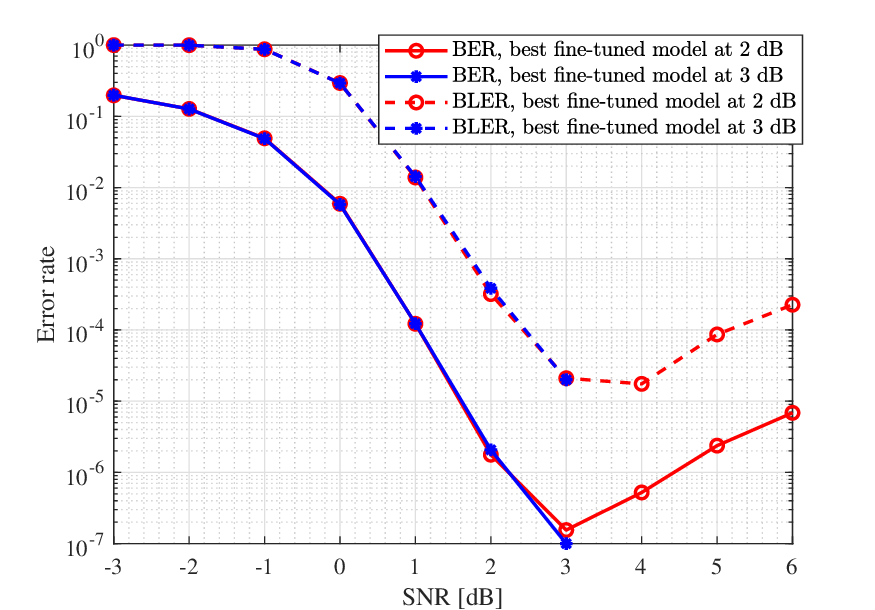}
	\caption{Error floor behavior of ProductAE $(24,20)\otimes(25,10)$, fine-tuned under $\gamma=1.25$ dB for the encoder training SNR. The blue-asterisk curves achieve the BER/BLER of zero at all reasonable SNRs of larger than $3$ dB, where $2\times 10^{6}$ codewords are used to test the error rate at each SNR point.}
	\label{efloor2}
\end{figure}

Fig. \ref{efloor1} shows the fine-tuning of ProductAE $(24,20)\otimes(25,10)$ under $\gamma=1.25$ dB for the encoder training SNR. Note that the decoder training SNR is randomly selected from the range $[\gamma-2.5,\gamma+1]=[-1.25,2.25]$ dB. Severe fluctuations are observed for the validation BER at $3$ dB (we will further discuss this fluctuation behavior toward the end of this section)\footnote{Note that a part of this fluctuation is as a result of not having large enough number of  validation samples at lower error rates. However, since we used $5\times10^{5}$ codewords, each composed of $k=200$ information bits, BERs larger than $10^{-6}$ have experienced at least $10$ bit errors.}. Due to this behavior, as seen in Fig. \ref{efloor2} (see red-circle curves), picking a model that has the best training performance at lower SNRs, e.g., at $2$ dB, does not guarantee a good performance at higher SNRs, e.g., at $3$ dB, and even may lead to an error floor at subsequent higher SNRs\footnote{Note that our testing of models with similar behaviors (to that of the red-circle curves in Fig. \ref{efloor2}) over larger ranges of SNRs show that the error rate curves actually experience an error floor rather than diverging to BER of 0.5 and BLER of 1. In other words, 
the error rates first increase a few degrees of magnitude and then
saturate at a higher flat BER/BLER.}. 

To resolve the error floor behavior in Fig. \ref{efloor2}, we note based on Fig. \ref{efloor1} that the validation BERs at lower SNRs, e.g., at $1$ and $2$ dB, are relatively flat, as the model observes large enough number of training samples from these lower SNRs during the fine-tuning. As such, the optimizer that minimizes the (surrogate of the) BER results in a reliable model over these lower SNRs (note that the saturation of the validation BERs at $1$ and $2$ dB means that the model is already converged). Therefore, one can reliably pick the model corresponding to the best (or a good) validation BER at the higher SNR of $3$ dB without tangibly degrading the performance at lower SNRs of $1$ and $2$ dB. Fig. \ref{efloor2} (see blue-asterisk curves) shows that this simple heuristic approach is able to completely eliminate the error floor at higher SNRs (both BER and BLER of zero were achieved for all feasible SNR values larger than $3$ dB, where $2\times10^6$ codewords were tested per SNR point) without affecting the performance at lower SNRs. We believe these kind of analysis (as well as fine-tuning models under larger/wider SNRs, as discussed in Section \ref{rob}) are essential  in generating models that avoid error floors at higher SNRs, especially in situations where the training BER/loss is fluctuating at some SNRs despite large enough number of validation samples.

\begin{figure}[t]
	\centering
	\includegraphics[trim=0.3cm 0.1cm 0 0,width=3.6in]{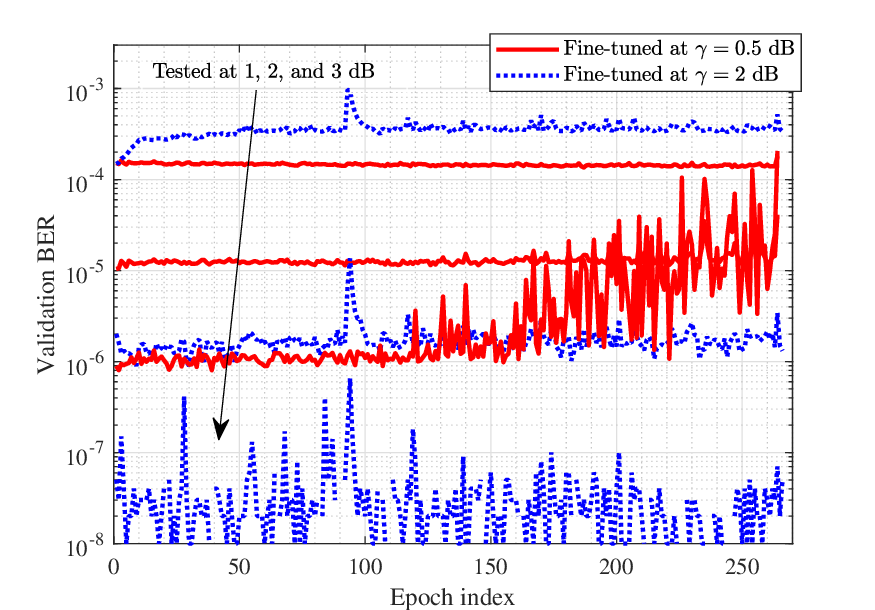}
	\caption{Fine-tuning of ProductAE $(24,20)\otimes(25,10)$ under two different values of $\gamma=0.5$ and $2$ dB for the encoder training SNR. Picking relatively small training SNRs not only degrades the performance at higher SNRs but can also cause the model to diverge at high SNRs. $5\times 10^{5}$ codewords are used for the validation of the BER across epochs.}
	\label{efloor3}
		\vspace{-0.25cm}
\end{figure}

The above experiments further confirm our earlier claim that \textit{the saturation of the performance is as a result of picking a model that is optimized for a given (range of) SNR but is not good for the subsequent larger values of SNR}. To better understand the aforementioned fluctuation behavior and realize why the models in some epochs can have worse performance at higher SNRs compared to lower SNRs, in Fig. \ref{efloor3} we investigate the fine-tuning of ProductAE $(24,20)\otimes(25,10)$ under two different values of $\gamma=0.5$ and $\gamma=2$ dB for the encoder training SNR. As seen, the fine-tuned model under $\gamma=0.5$ dB becomes more and more oblivious to higher SNRs (e.g., the validation at $3$ dB and even at $2$ dB) as we keep training. However, when the ProductAE is fine-tuned with larger SNRs, i.e., under $\gamma=2$ dB, the model does not diverge at higher SNRs (note that the fluctuations of the validation BER at $3$ dB for very low BERs are due to insufficient validation samples).
Therefore, we conjecture that the fluctuation of the training loss at higher SNRs is because the model does not pay attention to optimizing the performance at higher SNRs. In other words, the training SNRs are selected from a range that does not include any (significant number of) noisy samples from larger channel SNRs. As a result, the trained models completely overlook the larger values of SNR while only trying to optimize the models for the ranges of SNRs considered in training.

\subsection{Several Additional Investigations}\label{app_additional}

\subsubsection{Single Decoder Training SNR versus a Range}\label{app_singleeDecSNR}
\begin{figure}[t]
	\centering
	\includegraphics[trim=0.3cm 0.1cm 0 0,width=3.6in]{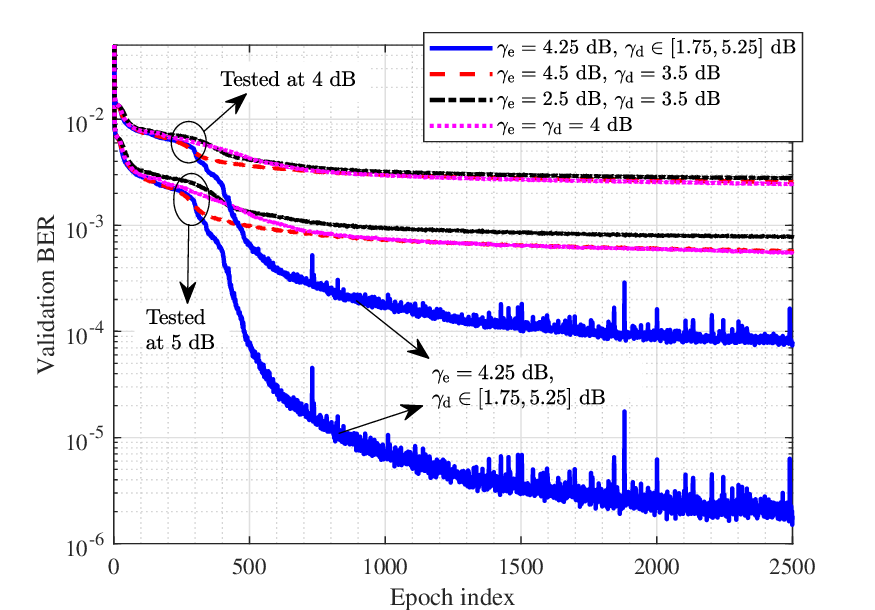}
\caption{ProductAE training under a range of SNR for the decoder training schedule is compared with training under a single  SNR point for the decoder. ProductAE $(24,20)\otimes(25,15)$ is trained under ${\rm lr_{enc}}={\rm lr_{dec}}=3\times 10^{-4}$.}
	\label{singleSNR}
		\vspace{-0.25cm}
\end{figure}

Fig. \ref{singleSNR} compares the training of the  ProductAE under a single decoder training SNR with that of our training approach, which uses a SNR range for the decoder training schedule. It is observed that using a range for the decoder training SNR results in a much better training across a wide range of testing SNRs. In other words, the single ProductAE model that is trained under a range of SNR for the decoder significantly outperforms the performance of multiple ProductAEs each specifically trained for a given single SNR point.

\subsubsection{Encoder Training with a Range of SNR}
As discussed in the paper, our ProductAE models are trained under a single SNR point (denoted by $\gamma$) for the encoder training schedule, and a range of SNR for the decoder training schedule. 
We also explored training the ProductAE encoder under a range for the channel SNR to generate the noise samples. However, our experiments showed an identical training performance to the case where a single SNR point was used for the encoder training schedule. This together with our investigations in Appendix \ref{app_singleeDecSNR}  can imply that the ProductAE encoding is a relatively simpler task compared to the ProductAE decoder training.

\begin{figure}[t]
	\centering
	\includegraphics[trim=0.3cm 0.1cm 0 0,width=3.6in]{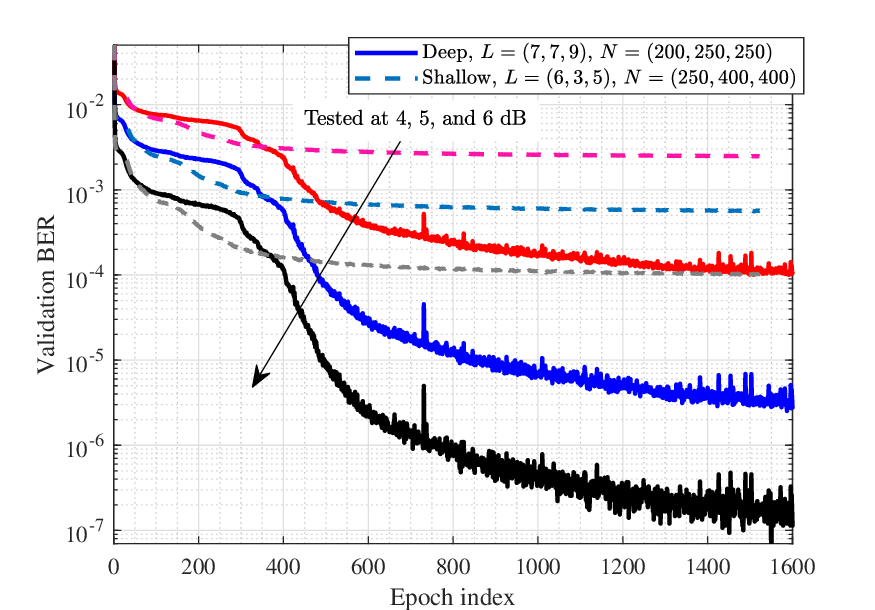}
	\caption{Comparison between deep and shallow architectures for the (decoder of) ProductAE $(24,20)\otimes(25,15)$. $\gamma_{\rm e}=4.25$ dB, $\gamma_{\rm d}\in[1.75,5.25]$ dB, and ${\rm lr_{enc}}={\rm lr_{dec}}=3\times 10^{-4}$ are used for the training. 
	}
	\label{lessdeep}
		\vspace{-0.2cm}
\end{figure}

\subsubsection{Deep Networks versus Shallow Networks} Given that the ProductAE decoder is composed of $2I$ pairs of neural networks, one may wonder whether the ProductAE architecture suffers from having very deep networks. Fig. \ref{lessdeep} compares the training of a relatively deep decoder architecture (the considered architecture throughout the paper) for the ProductAE $(24,20)\otimes(25,15)$ with that of a shallow decoder architecture. In the shallow architecture, if the number of layers is decreased by a factor of $\beta$ compared to the deep architecture, the number of nodes is increased by at least a factor of $\sqrt{\beta}$. This is to ensure that the total number of learnable parameters in the shallow architecture is at least as large as that of the deep architecture. Significant improvements are observed for the training of the deeper network, implying that not only our ProductAE architecture does not suffer from being very deep but also it can benefit from having neural networks with larger number of hidden layers.

\subsubsection{Neural Subtraction of Soft Information}
We investigated replacing the subtraction operation between the soft input and output of NN decoders (see Section \ref{subtrct}) with some FCNNs. 
To do so, we defined distinct FCNNs of input size $2$ and output size $1$, each doing an element-wise operation between an element of the soft input and an element of the soft output to substitute their difference.
Our extensive experiments, however, 
did not show improvements on the training performance through this modification.

\subsubsection{Regularizer}
We explored adding an $l_2$ regularizer to the loss function during the encoder training schedule.  Through extensive experiments, however, we did not observe an improvement upon our best training results.

\subsubsection{Power Normalization After the First Encoding NN}
Note that the outputs of the first neural encoder in our proposed 2D ProductAE are real-valued sequences. We investigated adding the power normalization block (similar to Eq. \eqref{norm}) after the first encoder, before feeding the data to the second encoder. Our experiments, however, showed that this modification  (slightly) improves the training, i.e., reduces the validation loss, only at initial epochs and does not change the performance once the models are converged. Therefore, this addition is observed to neither improve nor degrade the training, and thus can be considered optional in the ProductAE architecture.

\subsubsection{Random Batch Training} In the context of training models for error-correction coding, we have the flexibility of generating infinitely large number of training samples. This is because 1) for a given code dimension $k$, there are $2^k$ possible binary sequences to generate as input data; and 2) for each of these $2^k$ information sequences, one can generate infinitely many noise vectors to generate the corrupted sequences fed to the decoder. To take advantage of the availability of excessive number of training samples, we explored two options for generating the samples during the training: 1) generating a random batch of information sequences per training iteration of each training epoch; and 2) using the same batch of information sequences for all training iterations of each given epoch while randomly changing the data batch across the epochs. In both cases, the noise vectors were randomly generated for each input data. Our training experiments showed slightly faster convergence speed for the first approach with a slightly better performance (after convergence) for the second approach.
 We believe this is because each approach has its own advantages and disadvantages
  Specifically, the first approach helps the model to faster generalize by observing more diverse information sequences per training epoch, while the second approach helps the model to train better by focusing on the same fixed batch of information sequences for all training iterations at each epoch.

\bibliographystyle{IEEEtran}
\bibliography{IEEEabrv}
\end{document}